\documentclass[aps,11pt,nofootinbib, amsmath, amssymb, tightenlines]{revtex4-2}
\pdfoutput=1
\makeatletter

\def\@sect@ltx#1#2#3#4#5#6[#7]#8{%
    \@ifnum{#2>\c@secnumdepth}{%
        \def\H@svsec{\phantomsection}%
        \let\@svsec\@empty
    }{%
        \H@refstepcounter{#1}%
        \def\H@svsec{%
            \phantomsection
        }%
        \protected@edef\@svsec{{#1}}%
        \@ifundefined{@#1cntformat}{%
            \prepdef\@svsec\@seccntformat
        }{%
            \expandafter\prepdef
            \expandafter\@svsec
            \csname @#1cntformat\endcsname
        }%
    }%
    \@tempskipa #5\relax
    \@ifdim{\@tempskipa>\z@}{%
        \begingroup
        \interlinepenalty \@M
        #6{%
            \@ifundefined{@hangfrom@#1}{\@hang@from}{\csname @hangfrom@#1\endcsname}%
            {\hskip#3\relax\H@svsec}{\@svsec}{#8}%
        }%
        \@@par
        \endgroup
        \@ifundefined{#1mark}{\@gobble}{\csname #1mark\endcsname}{#7}%
        \addcontentsline{toc}{#1}{%
            \@ifnum{#2>\c@secnumdepth}{%
                \protect\numberline{}%
            }{%
                \protect\numberline{\csname the#1\endcsname}%
            }%
            #7}
    }{%
        \def\@svsechd{%
            #6{%
                \@ifundefined{@runin@to@#1}{\@runin@to}{\csname @runin@to@#1\endcsname}%
                {\hskip#3\relax\H@svsec}{\@svsec}{#8}%
            }%
            \@ifundefined{#1mark}{\@gobble}{\csname #1mark\endcsname}{#7}%
            \addcontentsline{toc}{#1}{%
                \@ifnum{#2>\c@secnumdepth}{%
                    \protect\numberline{}%
                }{%
                    \protect\numberline{\csname the#1\endcsname}%
                }%
                #8}%
        }%
    }%
    \@xsect{#5}}%

\makeatother

\usepackage[utf8]{inputenc}
\usepackage[T1]{fontenc} 
\usepackage{lmodern} 
\usepackage{graphicx}
\usepackage[margin=1in]{geometry}
\usepackage[all]{nowidow}

\usepackage{braket}
\usepackage{float}
\usepackage{bm}

\newcommand*{\PBH}{\mathrm{PBH}}
\newcommand*{\CDM}{\mathrm{CDM}}
\newcommand*{\diffd}{\mathrm{d}} 
\newcommand*{\ddv}[2]{\frac{\diffd #1}{\diffd #2}} 

\newcommand*{\Nusr}{N_{\rm USR}}
\newcommand*{\Mpl}{M_{\rm Pl}}
\newcommand*{\fpbh}{f_{\rm PBH}}
\newcommand*{\calP}{{\cal P}}
\newcommand*{\calR}{{\cal R}}
\newcommand*{\xmin}{x_{\rm min}}
\newcommand*{\xmax}{x_{\rm max}}
\newcommand*{\Ppeak}{{\mathcal P}_{\rm peak}}

\usepackage[linktocpage=true,colorlinks=true,linkcolor=blue,citecolor=blue,urlcolor=blue]{hyperref}
\hypersetup{pdftitle={Primordial black holes from single-field inflation: a fine-tuning audit},pdfauthor={4 authors}}

\usepackage[capitalise]{cleveref}

\begin{document}

\author{Philippa S.~Cole$^1$}
\email{p.s.cole@uva.nl}
\author{Andrew D.~Gow$^{2,3}$}
\email{andrew.gow@port.ac.uk}
\author{Christian T.~Byrnes$^3$}
\email{c.byrnes@sussex.ac.uk}
\author{Subodh P. Patil$^4$}
\email{patil@lorentz.leidenuniv.nl}

\affiliation{\\1) GRAPPA Institute, \\\mbox{University of Amsterdam, 1098 XH Amsterdam, The Netherlands}\\}

\affiliation{\\2) Institute of Cosmology and Gravitation, \mbox{University of Portsmouth, Portsmouth, PO1 3FX, United Kingdom}\\}

\affiliation{\\3) Department of Physics and Astronomy,\\University of Sussex, Brighton BN1 9QH, United Kingdom\\} 

\affiliation{\\4) Instituut-Lorentz for Theoretical Physics,\\Leiden University, 2333 CA Leiden, The Netherlands\\}

\date{16/08/2023}

\title{Primordial black holes from single-field inflation: a fine-tuning audit}

\begin{abstract}
All single-field inflationary models invoke varying degrees of tuning in order to account for cosmological observations. Mechanisms that generate primordial black holes (PBHs) from enhancement of primordial power at small scales posit inflationary potentials that transiently break scale invariance and possibly adiabaticity over a range of modes. This requires additional tuning on top of that required to account for observations at scales probed by cosmic microwave background (CMB) anisotropies. In this paper we study the parametric dependence of various single-field models of inflation that enhance power at small scales and quantify the degree to which coefficients in the model construction have to be tuned in order for certain observables to lie within specified ranges. We find significant tuning: changing the parameters of the potentials by between one part in a hundred and one part in $10^8$ (depending on the model) is enough to change the power spectrum peak amplitude by an order one factor. The fine-tuning of the PBH abundance is larger still by 1--2 orders of magnitude. We highlight the challenges imposed by this tuning on any given model construction. Furthermore, polynomial potentials appear to require significant additional fine-tuning to also match the CMB observations.
\end{abstract}

\maketitle

\tableofcontents

\section{Preliminaries}

Primordial black holes could have formed from the direct collapse of large amplitude density perturbations, if the inflationary process generated a large peak in the power spectrum on small scales~\cite{Zeldovich:1969ff,Garcia_Bellido_1996,Yokoyama:1995ex,Ivanov_1998,Green:2020_Primordial,Escriva:2022duf}.
It is straightforward to arbitrarily tune inflaton potentials to the degree required to realise any desired field dynamics at the classical level. It is less straightforward to do so at the quantum level. Given that it is necessarily the quantum effective potential that inflates and not its classical counterpart that would eventually quantum correct to it, this tuning must be imposed on the former, and is typically accomplished by invoking an approximate shift symmetry. As a corollary, correlation functions inherit only a weakly logarithmic scale dependence. If one seeks to concoct dynamics that result in anything other than this logarithmic scale dependence for correlation functions, additional tuning must be imposed on the effective potential.

It is well established that the classical (\textit{i.e.}~tree level) potential may bear little qualitative resemblance to the quantum effective potential. This is because the former represents merely the zeroth order in $\hbar$ bootstrap to the latter\footnote{In practice, calculating the exact quantum effective action is neither possible nor necessary depending on the quantities we're interested in calculating, with most tractable applications demanding only local corrections up to some finite order.}, from which all physical observables ultimately derive\footnote{This is vividly illustrated in the context of the Standard Model of particle physics, where the Higgs potential looks like a quartic potential around the electroweak vacuum with a tree level quartic coupling $\lambda \sim 0.13$. Renormalisation group improving the effective potential to next to next to next to leading order, however, results in a scale dependent quartic coupling that runs to negative values ($\lambda < 0$) at energy scales anywhere between $10^{10}$ GeV up to $10^{18}$ GeV~\cite{Buttazzo:2013uya} rendering the potential unbounded from below (the range corresponds to varying the top quark pole mass determined at low energies within its three sigma confidence interval).}. Paraphrased in the context of inflationary cosmology: it is straightforward to write down potentials that can account for large scale cosmological observations at the classical level with sufficiently tuned parameters. However, requiring inflation to last a sufficient number of e-folds requires that the inflaton undergo a sufficient excursion in field space, where the flatness of the potential is approximately preserved over the entirety of the excursion. This is much harder to arrange, and in the context of canonical single-field inflation, power counting implies a sensitivity to Planck suppressed operators up to mass dimension six, otherwise known as the eta problem (see \textit{e.g.}~\cite{Baumann:2009ds} for a review).

Imposing additional demands on inflation beyond requiring that it account for observations at the largest scales necessarily requires further tuning, the significance of which should not be dismissed. In the context of effectively single-field (although not necessarily single-clock) inflation~\cite{Byrnes:2019_Steepest}, models in which PBHs are formed by enhancements of small scale power over a specified range of comoving scales typically invoke some mechanism that results in the second parameter of the Hubble hierarchy, $\eta_H \equiv \dot{\epsilon}_H/(\epsilon_HH)$ to be negative for a sufficiently sustained period~\cite{Ivanov:1994pa,Kinney:2005vj,Garcia-Bellido:2017mdw,Dimopoulos:2017ged,Motohashi:2017kbs,Kannike:2017bxn,Liu:2020oqe,Karam:2022nym} where $\epsilon_H\equiv-\dot{H}/H^2$. Rather than being a parameter whose time dependence can be freely specified, $\epsilon_H$ is a `Wilson function' in the effective theory of the fluctuations (whether adiabatic or otherwise~\cite{Cheung:2007st,Senatore:2010wk, Achucarro:2012sm, Chluba:2015bqa}). The latter is obtained by perturbing the effective action for the \textit{background} inflaton around a consistent solution: one that minimises the \textit{background effective action}. Given that the latter is the product of a parametrically controlled derivative expansion, this translates into constraints on the time variation of the parameters determining the evolution of the perturbations\footnote{\label{fn:Rapid-end}See for instance, the discussion in appendix C of~\cite{Byrnes:2019_Steepest} regarding the quickest possible end to inflation.}, which to leading order in the adiabatic context can be taken as $\epsilon_H$ and the sound speed $c_s$ \cite{Chluba:2015bqa}. A straightforward corollary is that calculations of correlation functions obtained via matching calculations between phases that jump between different values of a given parameter (such as $\eta_H$) should be viewed with caution~\cite{Cole:2022xqc}, as these correspond to step function jumps in the time dependence of the parameter in question. Even if one were to take such discrete jumps merely as the limiting case of a very rapid transition, these cannot be made arbitrarily rapid without the introduction of additional hierarchies that would be challenging to realise at the level of the effective action. 
Simply put, were one to write down a tree level potential for the background that might effect a sudden transition, quantum corrections will smooth this transition out\footnote{Moreover, no matter how rapid one tries to transition in coordinate time by ignoring this caution and introducing the required hierarchy, large gradients cost expansion, and therefore e-folds which is what imprints on correlation functions (\cref{fn:Rapid-end} \textit{ibid}.).}, if not linearly interpolate completely in the absence of other relevant degrees of freedom (\textit{cf}.~\cref{fn:Convex}).  Similarly, in the context of phase transitions, it is unnatural to posit that they can be made arbitrarily sudden, and typically last an order of an e-fold\footnote{A distinction that was not lost on the authors of \cite{Felder:1998vq} in the related context of preheating, where `instant' means the order of an e-fold.}. Moreover, one must take care to factor in the transient non-adiabaticity that necessarily accompanies such transitions, typically neglected in single-field analyses.

In the context of single-field inflation, PBH formation via enhanced density perturbations requires the background inflaton field to decelerate ($\eta_H < 0$), with onset of rapid growth whenever one is no longer in the single-clock regime ($\eta_H < -3$).\footnote{For recent reviews of PBHs see \textit{e.g.}~\cite{Sasaki:2018_Primordial,Carr:2020xqk,Green:2020_Primordial,Escriva:2022duf,Bhattacharya:2023ztw}.} This is typically accomplished by demanding that the potential approach either an inflection point, or have some other feature that sufficiently decelerates the inflaton field roughly when the comoving scales of interest are exiting the Hubble radius, which, moreover, has to be done in a manner that is consistent with CMB constraints at large scales. Although one might envisage designing a variety of potentials at the classical level to obtain a peak for the PBH mass function at any given mass scale, doing so in the context of a realistic model construction imposes a variety of restrictions or additional factors that must be accounted for.

Since the publication of~\cite{Byrnes:2019_Steepest}, where it was shown under certain assumptions that the fastest the primordial power spectrum can grow as a function of comoving wavenumber is $k^4$, examples of inflationary potentials that enable the primordial power spectrum to grow even faster have been found. In a followup investigation~\cite{Cole:2022xqc}, it was shown that nevertheless, the super-$k^4$ growth does not translate to an effect on the distribution or abundance of primordial black holes that will be produced, especially once unrealistic arbitrarily rapid transitions have been smoothed. In this paper, we go on to quantify the degree of fine-tuning required in order for a given inflationary potential to result in rapid growth of the power spectrum over a range of comoving scales. The degree of fine-tuning required is significant in all of the cases we examine, drawn from three representative classes of single-field inflationary models, each having their own issues in addition to the fine-tuning. We find that either a feature must be incorporated that is implausible to realise at the level of the effective potential without considering additional relevant degrees of freedom, or, for more straightforward to realise features, either large-scale constraints on the tensor-to-scalar ratio are difficult to satisfy and/or the model acquires a pronounced sensitivity to initial conditions. We argue that producing a large peak in the primordial power spectrum from a plausible model construction of effectively single-field inflation obeying large-scale constraints on the tensor-to-scalar ratio poses a significant model building challenge that has yet to be satisfactorily met.

The outline of the paper is as follows: first we detail the various representative classes of single-field models that we focus on in this investigation and the reasons for choosing them. After this, we quantify the amount of fine-tuning required in each case on top of the requirements of successful inflation that matches large-scale observations, propagating this on to the parameters relevant for cosmological and astrophysical observations. For the purpose of quantifying the notion of fine-tuning, we adopt a measure of tuning proposed by~\cite{Azhar:2018lzd}, taking care to highlight the inevitable epistemological shortcomings of any particular choice of measure. See \textit{e.g.}~\cite{Hertzberg:2017dkh,Nakama:2018utx,Carr:2019hud,Braglia:2022phb,Qin:2023lgo,Animali_2023} for studies of inflationary potential fine-tuning in the PBH context. We conclude by discussing the ramifications of our findings in the context of realistic model constructions and what new ingredients may be required. We defer various technical details to the appendices. 

In what follows, we work in units where $c=\hbar=1$ and $\Mpl^2=1/(8\pi G)=1$.

\section{Prototype potentials}

Instead of aiming for comprehensive coverage of the various models discussed in the literature, we select four examples for the purposes of this investigation. These belong to three prototypical classes in which either a particular functional form for the potential, or a particular field dynamic mechanism can be identified upon which other models represent variations upon a theme. We list these representative models below and detail their advantages and drawbacks from various perspectives before addressing the parametric tuning required in each case. In restricting ourselves to the canonical single-field context, our investigation is by no means exhaustive as an audit into fine-tuning issues for PBH production in inflation in general\footnote{Moreover, even within the context of effectively single-field inflation, we do not audit models with varying speeds of sound~\cite{KAMENSHCHIK2019201,romano2020sound,Cai_2018} where sufficiently rapid variations of $c_s$ generate PBHs through parametric resonance. Premised as effectively single-field models, the variations required to enhance small scale power to produce any significant amount of PBHs are of such rapidity as to violate the validity of the effectively single-field description~\cite{Achucarro:2012sm}.}. For instance, much work has recently been done on PBH production in multi-field inflation~\cite{Palma:2020_Seeding,Braglia:2020eai,Braglia:2020taf,Fumagalli:2020_Turning,Ahmed:2021ucx, Geller:2022nkr,Iacconi_2022, Kawai:2022emp,Braglia:2022phb,Qin:2023lgo,Ozsoy:2023ryl}, and it would be of interest to undertake a similar investigation in that context.  

\begin{itemize}
	
	\item {\bf{Deceleration via superposed feature: }} The model we focus on is that presented in Mishra and Sahni~\cite{Mishra:2019pzq} (although others including~\cite{ZhengRuiFeng:2021zoz,Inomata:2021tpx,Frolovsky_2022} invoke a similar dynamic) where the required deceleration is obtained through overshooting the minimum of a Gaussian bump in the potential that leads to ultra-slow-roll with single-clock slow roll on either side. The period of single-clock slow roll prior to the feature is arranged so that large scale CMB constraints are satisfied, whereas the period that follows is less constrained, although constraints on small scale power from other tracers are also satisfied~\cite{Byrnes:2019_Steepest,Inomata:2018epa,Dalianis:2018ymb,Kalaja:2019_From}. The localised feature is added by hand on top of a potential that would otherwise sustain single-clock slow roll. Tuning the shape and location of the feature allows one to match large scale observations and generate a peak for the PBH mass function at any desired scale\footnote{However, depending on the desired peak for the PBH mass function, one does impact the CMB observables by the lead in deceleration even if this scale corresponding to the peak exits the Hubble radius far beyond CMB scales.}.
	
	Aside from the \textit{ad hoc} nature of the added potential feature in this class of models, it cannot be realised in isolation without having to account for additional ingredients in some way -- a caveat that is relevant to all the examples considered in this paper. The reason for this is that the true \textit{vacuum} effective potential is necessarily convex and cannot admit a feature\footnote{\label{fn:Convex}Although the convexity of the effective potential is standard textbook physics (\textit{cf}.~Chapter 11.3 of~\cite{Peskin:1995ev}), this may come as a surprise to some readers used to seeing all manners of potentials in the cosmology literature. Simple arguments as to why this is so can be found in~\cite{Haymaker:1983xk} and~\cite{Amer:1982ri} for the single-field and multi-field cases, respectively.}. The additional ingredients required to generate a feature could for example take the form of non-adiabatic driving by some other classical source field, additional degrees of freedom coupled to the inflaton that start to propagate at a particular energy scale (and so affect the effective potential through threshold effects~\cite{Casas:1998cf}), or even background moduli fields that are not heavy enough to permit truncation as will be the case for most of the examples considered further. Accounting for these additional degrees of freedom even within the effectively single-field context places restrictions on the form of any feature one would like to generate \cite{Chluba:2015bqa, Achucarro:2012fd}, and more generally may necessitate accounting for additional (isocurvature) interactions that could qualitatively alter one's conclusions.

	\item {\bf Deceleration via polynomial potential feature: } Inflection points are more naturally realised at the level of the effective potential. They can arise for instance via renormalisation group improving the potential in an effective theory~\cite{Isidori:2007vm} when one matches across thresholds corresponding to the mass of a particle that starts to propagate below the threshold~\cite{Casas:1998cf}\footnote{Inflection points can also arise from the behaviour of logarithmic factors that can be relevant when considering RG running \textit{between} any two thresholds. However this running derives from energy differences alone, for which field values are only a proxy. Once renormalization conditions are fixed at the CMB pivot scale, the effects of the logs can only be significant if one looks at modes which exit the horizon when the potential energy has dropped significantly, which is not the case for the classes studied in our investigation. This is in contrast with Higgs inflation, where renormalization conditions are imposed at the mass of the Z-boson and run up to the scale of inflation -- a running over many orders of magnitude (cf. footnote \ref{HIRG}).}. They are efficient in enhancing power over a limited range of small scales, however, inflection points in simple polynomial potentials struggle to produce large scale spectra which match observations. 
	   
	The example of a cubic potential has the advantage of being the ``simplest'' possible model, having only one free parameter (the coefficient of the cubic term with suitable choice for the origin in field space) in addition to the overall scaling which sets the amplitude of the power spectrum on scales which exit long before reaching the inflection point. Unfortunately, we found this simple model generically suffers from several issues, including the tensor-to-scalar ratio being far too large, the power spectrum peak being an extended plateau, and inflation only ending very long after traversing the inflection point which requires an additional \textit{ad hoc} feature to be added in order to end inflation so that the power spectrum peak occurs on scales smaller than those corresponding to CMB scales.
	
	Hertzberg and Yamada~\cite{Hertzberg:2017dkh} have found a way to flatten the potential on CMB scales by tuning a quintic potential to have two flat sections, one of which generates the CMB scales (with a small enough tensor-to-scalar ratio) and one with a local minimum and maximum with tiny amplitudes, which generates the peak required for PBH production. Apart from the additional fine-tuning this requires of the potential, their model also requires the initial conditions to be fine-tuned such that the inflaton starts with zero or very small kinetic energy in the flat regions which generates the CMB scales, and fails when the initial conditions are outside a narrow range.
	
	\item {\bf Deceleration via non-polynomial potential feature: } We consider here two examples of potentials that flatten and have a feature induced through non-polynomial factors, see also \textit{e.g.}~\cite{Ragavendra:2020sop,Bhaumik_2020,Gangopadhyay_2022}. The first is drawn from the inflection-point model of Germani and Prokopec~\cite{Germani17}, and is inspired from models of Higgs inflation where the non-minimal coupling term for the Higgs $\mathcal L \sim \xi H^\dag H R$ (where $R$ is the Ricci scalar) is modelled for the singlet sector as a scalar-tensor coupling of the form $\mathcal L \sim \xi \phi^2 R$, which upon transforming to the Einstein frame rescales the original Jordan frame potential as\footnote{\label{fn:Canonical-kinetic}Although one can certainly consider this rescaled potential on its own merits as was done in~\cite{Germani17}, the conformal transformation would also rescale the kinetic term in any realistic model construction. Making a field redefinition to canonically normalised field variables in this context will also result in a potential with exponential characters, as per the second example studied in~\cite{Cicoli:2018asa}.} $V(\phi) \to V(\phi)/(1 + 2\xi\phi^2)^2$. The second is grounded in a string-theoretic construction presented in Cicoli \textit{et al}.~\cite{Cicoli:2018asa}. Here, the potential is given by sums of exponential characters, which, moreover, are flattened via additional exponential factors arising from field and frame redefinitions so that the inflaton and the graviton are canonically normalised and have no kinetic mixing. Furthermore, the effective potential is itself constructed to next to next to leading order in loop corrections, and is therefore arguably the most parametrically under control. However, the authors of this work warn that the spectral index is 2--3 sigma too low compared to the observed CMB value at the pivot scale $k=0.05\,{\rm Mpc^{-1}}$. This model also has a very small amplitude local minimum and maximum feature which is responsible for PBH generation\footnote{\label{HIRG}On the other hand, the authors of \cite{Ezquiaga:2017fvi} proposed a model that generated PBHs consistent with large scale observations via an inflection point in the context of Higgs inflation, where significant RG running is induced by large energy excursions without crossing any thresholds (see also \cite{Ballesteros:2020qam,Ballesteros:2017fsr} in a more general context). A more complete survey would certainly have to extend to this class of models.}.
	
\end{itemize}

\section{Parametric sensitivity of prototype potentials}
\label{sec:Parametric-sensitivity}
In what follows, we present the power spectra for each of the prototypical potentials discussed in the previous section, highlighting the parametric sensitivity of the power spectrum amplitude to the potential parameters in each case. In the following section, we quantify the corresponding degree of fine-tuning. For each of the examples presented below, we define $\phi_0$ as the initial field value, and $\phi_{\rm CMB}$ as the field value at the CMB pivot scale for a fiducial set of parameter values, which is chosen in a unique way for each potential, as discussed in the following sections. The fiducial sets do not match those in the original references because we have chosen them such that the peak in the power spectrum grows from the amplitude measured at the CMB pivot scale to $\mathcal{P}\sim 5 \times 10^{-3}$ in all cases \cite{Gow:2021_ACPS}, on a scale corresponding to asteroid-mass PBHs. Choosing the same peak amplitude ensures a fairer comparison of the fine-tuning between models. Each of the plots display the largest possible shift in the model's most fine-tuned parameter that doesn't lead to the power spectrum becoming larger than unity (usually due to the inflaton remaining too long in the flat section of the potential, or getting stuck there forever). We shift the parameters in increments of $10^{-n}$ for integer $n$ relative to the fiducial values. We also plot the power spectra for the same shift but with the opposite sign such that the peak amplitude is decreased. Note that larger shifts in this direction aren't disallowed in the same way as an excessive growth of the power spectrum, but small power spectrum peaks, $\mathcal{P}<\mathcal{O}(10^{-3})$, will lead to negligible PBH production.

\subsection{Deceleration via superposed feature}

The potential presented in~\cite{Mishra:2019pzq} is given by
\begin{equation}
V(\phi) = V_0\frac{\phi^{n}}{\phi^n + M^n}\left(1 + A\exp\left[-\frac{(\phi - \phi_d)^2}{2\sigma^2}\right]\right),
\label{eq:Mishra-potential}
\end{equation}
with parameters $A$, $\phi_d$ and $\sigma$ characterising the height, position and width of the Gaussian bump added to an otherwise slow-roll inflationary potential described by the parameters $V_0$, $M$ and $n$. Our choice of fiducial values for the potential that lead to a power spectrum peak amplitude of approximately $5\times10^{-3}$ are given in \cref{tab:Mishra-fiducial}. For this potential, the value $\phi_\text{CMB} = 3$ is stated in~\cite{Mishra:2019pzq}.
\begin{table}[H]
\caption{Fiducial parameters for the potential \cref{eq:Mishra-potential}. Recall that we have set $\Mpl=1$.}
\label{tab:Mishra-fiducial}
\centering
\begin{tabular}{ccccccc}
	 $\bm{\phi_0}$ & $\bm{\phi_{\rm CMB}}$ & $\bm{n}$ & $\bm{M}$ & $\bm{A}$ & $\bm{\sigma}$ & $\bm{\phi_d}$\\ \hline
	 4 & 3 & 2 & $1/2$ & $1.17 \times 10^{-3}$ & $1.59 \times 10^{-2}$ & 2.18812
\end{tabular}
\end{table}

\begin{figure}[H]
	\centering
	\includegraphics[width=0.8\textwidth]{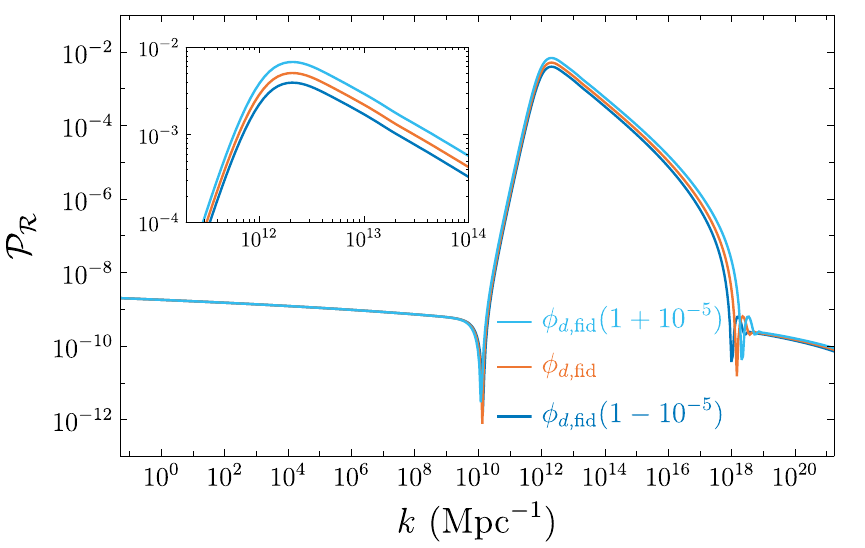}
	\caption{Power spectra for the potential \cref{eq:Mishra-potential}. Fiducial potential values correspond to the orange line, blue lines correspond to variations of the parameter $\phi_d$, which parametrises the position of the Gaussian bump, by a factor of $1\pm10^{-5}$.}
	\label{fig:Mishra-power-spectra}
\end{figure}

We consider shifts in the three parameters of the Gaussian bump, $A$, $\sigma$, and $\phi_d$. Of these three, the most fine-tuned is the bump position $\phi_d$. For this parameter, we show the power spectrum in \cref{fig:Mishra-power-spectra} for the fiducial value (orange), and for the largest shifts which do not make the power spectrum grow larger than unity (blue), which are $1\pm10^{-5}$. We take note of the generic fact that the enhancement of the power spectrum is not symmetric for a given perturbation of the fiducial value because of the non-linear nature of the map between the parameter of the potential and the peak amplitude of the power spectrum.

\subsection{Deceleration via polynomial potential feature}
\textit{A priori}, finite-order polynomial potentials might seem like the first place to look for a mechanism to decelerate the background inflaton via one or more inflection points. Arranging for an inflection point at a given point in field space to sufficiently enhance the primordial power spectrum is straightforward enough with a locally cubic or higher-order polynomial expansion. However, simultaneously satisfying large scale CMB constraints while the field is higher up the potential is challenging to the point that the only example in the literature known to us that accomplishes this in the Einstein frame also requires fine-tuning of the inflaton initial conditions (although see \textit{e.g.}~\cite{Ballesteros:2020qam} for a polynomial construction that becomes non-polynomial in the Einstein frame). 
As presented in~\cite{Hertzberg:2017dkh}, this example uses a quintic potential,
\begin{equation}
V(\phi) = c_0 + \frac{c_1}{\Lambda}\phi + \frac{c_2}{2\Lambda^2}\phi^2 + \frac{c_3}{3!\Lambda^3}\phi^3 + \frac{c_4}{4!\Lambda^4}\phi^4 + \frac{c_5}{5!\Lambda^5}\phi^5,
\label{eq:Hertzberg-potential}
\end{equation}
with our choice of fiducial coefficients $c_i$ of the fifth-order polynomial given in \cref{tab:Hertzberg-fiducial}. In this model, $\phi_\text{CMB}$ is defined to be the field value 30 e-folds before the beginning of the USR phase. For more details, see~\cite{Hertzberg:2017dkh}.
\begin{table}[H]
	\caption{Fiducial parameters for the potential \cref{eq:Hertzberg-potential}. Note that the sign difference in $c_3$ with respect to \cite{Hertzberg:2017dkh} is due to a typo in that paper.}
	\label{tab:Hertzberg-fiducial}
	\centering
	\begin{tabular}{ccccccccc}
		$\bm{\phi_0}$ & $\bm{\phi_{\rm CMB}}$ & $\bm{\Lambda}$ & $\bm{c_0}$ & $\bm{c_1}$ & $\bm{c_2}$ & $\bm{c_3}$ & $\bm{c_4}$ & $\bm{c_5}$ \\ \hline
		-$|\Lambda^3/c_3|$ & $5.51674\times10^{-4}$ & 0.3 & 1 & $(2\pi^2\Lambda^4)/(4225 c_3)$ & 0 & -0.52 & 1 & -0.640725043
	\end{tabular}
\end{table}

\begin{figure}[H]
	\centering
	\includegraphics[width=0.8\textwidth]{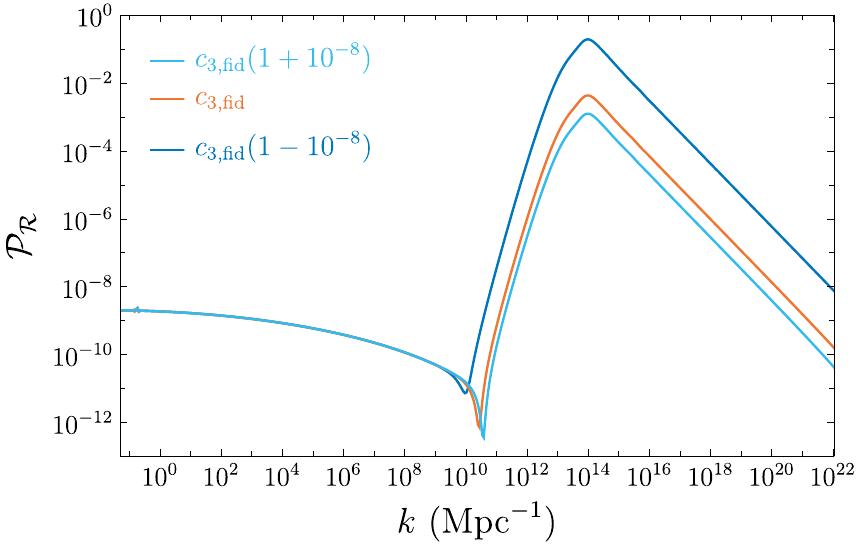}
	\caption{Power spectra for the potential \cref{eq:Hertzberg-potential}. Fiducial potential values correspond to the orange line, blue lines correspond to variations of the parameter $c_3$ by a factor of $1\pm10^{-8}$.}
	\label{fig:Hertzberg-power-spectra}
\end{figure}

We consider variations in $c_3$ and $c_5$, and find that $c_3$ is the most tuned, such that only changes to this parameter by a factor of up to $1\pm10^{-8}$ still allow for a large peak and successful inflation. The fiducial power spectrum is shown in orange in \cref{fig:Hertzberg-power-spectra}, and the shifted power spectra in blue. We note that~\cite{Hertzberg:2017dkh} demonstrated the extreme sensitivity of this model to variations in $c_5$, but we have found the potential is slightly more sensitive to variations in $c_3$. Of all the prototypical potentials that we've studied, this naïvely represents the largest parametric sensitivity. We note also that this particular potential has an additional sensitivity to the initial conditions of the inflaton field value in order to satisfy constraints on the tensor-to-scalar ratio on large scales. This is because this model is reliant on the field starting on a very flat region of the potential such that it is off-attractor, and so its trajectory, the duration of inflation, and the resulting power spectrum are all subject to large changes if one instead chooses an initial field value much higher up in the potential.
This jeopardises one of the key merits of the inflationary paradigm, namely that a successful period of inflation does not depend on a specific choice for the initial conditions, and must be considered as a necessary extra cost of polynomial potentials that produce PBHs and respect CMB constraints on large scales. Furthermore, the fact that $\Delta\phi/\Lambda\sim4$ over the field range of interest implies that the truncation of \cref{eq:Hertzberg-potential} to quintic order necessitates tuning an infinite number of parameters including and beyond $c_6$ to be very small or vanishing\footnote{Unless these are tamed by particular relations between all higher-order coefficients that permit their resummation to finite and/or logarithmically running quantities, which is non-trivial to realize.}. This tuning is aggravated by the fact that it has to satisfy renormalisation group running even if we were to set them to zero at any particular scale.

\subsection{Deceleration via non-polynomial potential feature}
Here we consider two examples of potentials that achieve the requisite deceleration of the inflaton field via field excursions that are large relative to the mass scale\footnote{Loosely, the cutoff.} that would ordinarily suppress higher-dimensional operators in the context of a polynomial expansion. That is, the shape of the potential is deformed in a manner that is not adequately captured by a truncation to a polynomial expansion. The first example accomplishes this with non-exponential factors (see however \cref{fn:Canonical-kinetic}), whereas the second does this with exponential factors. See also \textit{e.g.}~\cite{Ozsoy_2018,Ozsoy_2021}.

\subsubsection{Non-exponential character}

The potential presented in~\cite{Germani17} (adapted from \cite{Garcia-Bellido:2017mdw}) is given by
\begin{equation}
V(\phi) = \frac{\lambda}{12}\phi^2v^2\frac{6 - 4a\frac{\phi}{v} + 3\frac{\phi^2}{v^2}}{\left(1 + b\frac{\phi^2}{v^2}\right)^2},
\label{eq:Germani-potential}
\end{equation}
where an inflection point is generated for
\begin{equation}
b = 1 - \frac{a^2}{3} + \frac{a^2}{3}\left(\frac{9}{2a^2} - 1\right)^\frac{2}{3}. \label{eq:Germani-b}
\end{equation}
Our choice of fiducial parameter values is given in \cref{tab:Germani-fiducial}. In this case, $\phi_\text{CMB}$ is defined to be the field value 62 e-folds before the end of inflation.
\begin{table}[H]
	\caption{Fiducial parameters for the potential \cref{eq:Germani-potential}.}		\label{tab:Germani-fiducial}
	\centering
	\begin{tabular}{cccccc}
		$\bm{\phi_0}$ & $\bm{\phi_{\rm CMB}}$ & $\bm{a}$ & $\bm{b}$ & $\bm{\lambda}$ & $\bm{v}$ \\ \hline
		3 & 2.4719 & $1/\sqrt{2}$ & \cref{eq:Germani-b} & $1.86 \times 10^{-6}$ & 0.19669
	\end{tabular}
\end{table}

\begin{figure}[H]
	\centering
	\includegraphics[width=0.8\textwidth]{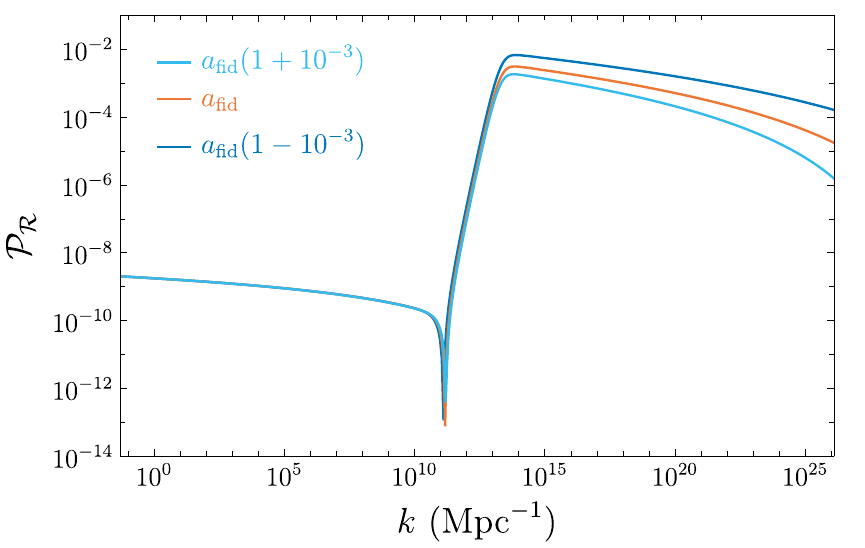}
	\caption{Power spectra for the potential \cref{eq:Germani-potential}. Fiducial potential values correspond to the orange line, blue lines correspond to variations of the parameter $v$ by a factor of $1\pm10^{-3}$. }
	\label{fig:Germani-power-spectra}
\end{figure}

We consider shifts in the parameters $a$ and $v$. The parameter $b$ is defined by $a$ through \cref{eq:Germani-b}, and $\lambda$ sets the overall scale (and hence the CMB normalisation), so it has no impact on relative changes in the power spectrum. We find that $a$ is the most fine-tuned parameter, for which we show the fiducial power spectrum in orange in \cref{fig:Germani-power-spectra}, and the power spectra for shifts in $a$ of $1\pm10^{-3}$ in blue. Of the potentials that we've studied, this example exhibits the least parametric sensitivity in terms of the effect on the peak amplitude of the power spectrum, although we note that a canonical field redefinition will increase the fine-tuning analogously to the potential of \cref{eq:Cicoli-potential}.

\subsubsection{Exponential character}

The potential presented in~\cite{Cicoli:2018asa} is given by
\begin{equation}
V(\phi) = V_0\left[C_1 - e^{-\frac{1}{\sqrt{3}}\hat{\phi}}\left(1 - \frac{C_6}{1 - C_7e^{-\frac{1}{\sqrt{3}}\hat{\phi}}}\right) + C_8e^{\frac{2}{\sqrt{3}}\hat{\phi}}\left(1 - \frac{C_9}{1 + C_{10}e^{\sqrt{3}\hat{\phi}}}\right)\right],
\label{eq:Cicoli-potential}
\end{equation}
where the potential parameters relate to the parameters of the underlying string construction that generates it as:
\begin{alignat*}{3}
	C_6 &= \frac{A_W}{C_W}, \quad &&C_7 = \frac{B_W}{\Braket{\tau_{K_3}}^{1/2}}, \quad &&C_8 = 0, \\
	C_8C_9 &= \frac{G_W}{\Braket{\mathcal{V}}}\frac{\Braket{\tau_{K_3}}^{3/2}}{C_W}, \quad &&C_{10} = \frac{R_W}{\Braket{\mathcal{V}}}\Braket{\tau_{K_3}}^{3/2}, &&
\end{alignat*}
with $C_1$ chosen such that $V_\text{min} = 0$. The fiducial values that we use are given in \cref{tab:Cicoli-fiducial}. Here, $\phi_\text{CMB}$ is defined to be the field value approximately 53 e-folds before the end of inflation, as shown in Fig.~3 of~\cite{Cicoli:2018asa}.
\begin{table}[H]
	\caption{Fiducial parameters for the potential \cref{eq:Cicoli-potential}.}		\label{tab:Cicoli-fiducial}
	\centering
	\begin{tabular}{cccccccc}
		$\bm{\phi_0}$ & $\bm{\phi_{\rm CMB}}$ & $\bm{A_W}$ & $\bm{B_W}$ & $\bm{C_W}$ & $\bm{\langle\tau_{K_3}\rangle}$ & $\bm{G_W/\langle\mathcal{V}\rangle}$ & $\bm{R_W/\langle\mathcal{V}\rangle}$ \\ \hline
		12 & 9.33731 & 2/100 & 1 & $4/100$ &14.30 &$3.08054\times10^{-5}$&$7.071067\times10^{-4}$
	\end{tabular}
\end{table}

\begin{figure}[H]
	\centering
	\includegraphics[width=0.8\textwidth]{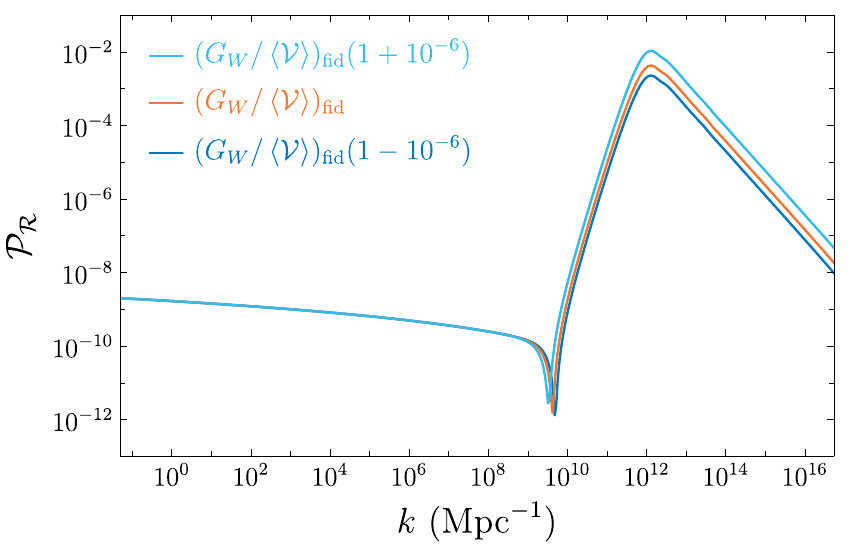}
	\caption{Power spectra for the potential \cref{eq:Cicoli-potential}. Fiducial potential values correspond to the orange line, blue lines correspond to variations of the parameter $G_W/\Braket{\mathcal{V}}$ by a factor of $1\pm10^{-6}$.}
	\label{fig:Cicoli-power-spectra}
\end{figure}

We consider variations in $G_W/\Braket{\mathcal{V}}$ and $R_W/\Braket{\mathcal{V}}$, and find that the first of these is the most fine-tuned. The power spectrum produced by this potential and our choice of fiducial values is shown by the orange line in \cref{fig:Cicoli-power-spectra} and the power spectra for shifts of $1\pm10^{-6}$ are shown in blue.

\section{From potential fine-tuning to tuning of PBH abundances}
\label{sec:Fine-tuning}

Each of the potentials studied above are a means to an end: the production of primordial black holes with a particular abundance and mass function. In this section we aim to quantify a measure of how finely tuned the peak amplitude of the power spectrum and the mass fraction of PBHs are relative to parameters of the underlying model construction that generated them. In order to do so, we have to choose a particular measure for this fine-tuning. No single choice is definitive or completely free of implicit priors, however the measure we choose can still be informative for the practical purpose of quantifying the degree of fine-tuning in a given model construction, and particularly when it comes to gauging the relative amount of tuning between any two examples. The measure we focus on was presented in~\cite{Azhar:2018lzd}, and corresponds to taking the logarithmic derivative of a given observable $\mathcal{O}$ with respect to the logarithm of any of the parameters $p$ it depends on either explicitly, or implicitly through intermediate quantities or convolutions\footnote{An intuitive way to understand this fine-tuning is that if the parameter $p$ varies by $\delta\ll1$ then  
$$\epsilon_{\Ppeak} \simeq  \frac{\ln(\Ppeak(p(1+\delta))/\Ppeak(p))}{\delta}.$$
Hence a fine-tuning value of $10^{4}$ means that changing $p$ by 0.01$\%$ leads to an order unity change in the observable. We critique and examine the relative merits of this measure in \cref{app:Fine-tuning-criteria}.},
\begin{equation}
	\epsilon_\mathcal{O} = \ddv{\log\mathcal{O}}{\log p}.
	\label{eq:Fine-tuning-parameter}
\end{equation}

For the superposed feature-overshoot potential \cref{eq:Mishra-potential}, we compute the fine-tuning measures around the fiducial parameter values of the Gaussian bump as laid out in \cref{tab:Mishra-fine-tuning}. We note that the fine-tuning only varies by a factor of 3 between the three parameters which model the Gaussian bump. This demonstrates that growth of the power spectrum (due to the duration of USR) is due to the combined height and width of the peak, and most importantly in this case the position of the peak, which is related to the initial field velocity before USR begins. Note that we define $\epsilon_{f_\PBH}$ and $\rho$ later in \cref{eq:pbh_tune,eq:pbh_rho}.
\begin{table}[H]
	\centering
	\caption{Fiducial and fine-tuning parameters for the potential \cref{eq:Mishra-potential}.}
	\label{tab:Mishra-fine-tuning}
	\begin{tabular}{ccrrc}
		& \textbf{Fiducial} & \multicolumn{1}{c}{$\bm{\epsilon_{\Ppeak}}$} & \multicolumn{1}{c}{$\bm{\epsilon_{f_\PBH}}$} & $\bm{\rho}$ \\ \hline
		$\bm{A}$ & $1.17\times10^{-3}$ & $8.9\times10^3$ & $2.0\times10^5$ & 23 \\
		$\bm{\sigma}$ & $1.59\times10^{-2}$ & $-8.1\times10^3$ & $-1.9\times10^5$ & 23 \\
		$\bm{\phi_d}$ & 2.18812 & $2.7\times10^4$ & $6.2\times10^5$ & 23 \\ \hline
	\end{tabular}
\end{table}
For the polynomial feature potential \cref{eq:Hertzberg-potential}, we compute the fine-tuning of the coefficients of the cubic and quintic terms in \cref{tab:Hertzberg-fine-tuning}.
\begin{table}[H]
	\centering
	\caption{Fiducial parameters and fine-tuning parameters for the potential  \cref{eq:Hertzberg-potential}.}
	\label{tab:Hertzberg-fine-tuning}
	\begin{tabular}{ccrrc}
		& \textbf{Fiducial} & \multicolumn{1}{c}{$\bm{\epsilon_{\Ppeak}}$} & \multicolumn{1}{c}{$\bm{\epsilon_{f_\PBH}}$} & $\bm{\rho}$ \\ \hline
		$\bm{c_3}$ & $-0.52$ & $-1.8\times10^8$ & $-4.7\times10^9$ & 27 \\
		$\bm{c_5}$ & $-0.640725043$ & $-1.7\times10^8$ & $-4.7\times10^9$ & 27 \\ \hline
	\end{tabular}
\end{table}

For the non-polynomial non-exponential inflection point potential \cref{eq:Germani-potential}, we show the sensitivity around the fiducial value for the most finely-tuned parameter of the potential \cref{eq:Germani-potential} in \cref{tab:Germani-fine-tuning}.
\begin{table}[H]
	\centering
	\caption{Fiducial and fine-tuning parameters for the potential \cref{eq:Germani-potential}.}
	\label{tab:Germani-fine-tuning}
	\begin{tabular}{ccrrc}
		& \textbf{Fiducial} & \multicolumn{1}{c}{$\bm{\epsilon_{\Ppeak}}$} & \multicolumn{1}{c}{$\bm{\epsilon_{f_\PBH}}$} & $\bm{\rho}$ \\ \hline
		$\bm{a}$ & $1/\sqrt{2}$ & $-6.0\times10^2$ & $-2.2\times10^4$ & 37 \\
		$\bm{v}$ & $0.19669$ & $4.4\times10^2$ & $1.6\times10^4$ & 37 \\ \hline
	\end{tabular}
\end{table}

Finally, we show the results for the two most finely tuned parameters for the exponential feature potential \cref{eq:Cicoli-potential} in \cref{tab:Cicoli-fine-tuning}.
 
\begin{table}[H]
	\centering
	\caption{Fiducial and fine-tuning parameters for the potential \cref{eq:Cicoli-potential}.}
	\label{tab:Cicoli-fine-tuning}
	\begin{tabular}{ccrrc}
		& \textbf{Fiducial} & \multicolumn{1}{c}{$\bm{\epsilon_{\Ppeak}}$} & \multicolumn{1}{c}{$\bm{\epsilon_{f_\PBH}}$} & $\bm{\rho}$ \\ \hline
		$\bm{G_W/\langle\mathcal{V}\rangle}$ & $3.08054\times10^{-5}$ & $7.5\times10^5$ & $2.2\times10^7$ & 29 \\
		$\bm{R_W/\langle\mathcal{V}\rangle}$ & $7.071067\times10^{-4}$ & $-6.8\times10^5$ & $-2.0\times10^7$ & 29 \\ \hline
	\end{tabular}
\end{table}
Although it may seem that the exponential character model of~\cite{Cicoli:2018asa} is much more finely tuned than that of~\cite{Germani17}, we stress that this may turn out to be an artificial distinction in a more realistic analysis, highlighting one of the caveats of conducting an audit with a particular choice of measure. The reason for this (stressed in \cref{fn:Canonical-kinetic}) comes down to comparing apples to apples: although one is entitled to take the potential \cref{eq:Germani-potential} on face value as being accompanied by a canonically normalised kinetic term, such potentials typically arise when one has made a conformal transformation from a frame where the scalar field in question was non-minimally coupled to one where it is minimally coupled (as motivated in~\cite{Germani17}). In the minimally coupled frame, the field kinetic term will not be canonically normalised, and so a non-linear field redefinition remains to be performed (as implemented in~\cite{Cicoli:2018asa}). Consequently, the parameters of the potential \cref{eq:Germani-potential} will also be subject to this non-linear transformation (which is exponential in the context of Higgs inflation), and re-evaluating the fine-tuning measure for the potential in the canonically normalised field basis will diminish the difference naïvely inferred from \cref{tab:Germani-fine-tuning,tab:Cicoli-fine-tuning}.

The fine-tuning parameter $\epsilon_{\Ppeak}$ is a measure of the sensitivity of the peak amplitude of the primordial power spectrum to small changes in various parameters of the potential. Inferring the amplitude of the power spectrum from observations could be done via observing the stochastic gravitational wave background, whose amplitude is set by the square of the power spectrum amplitude~\cite{Domenech:2021ztg}. Alternatively, (non-)observations of the abundance of PBHs could also be used. In this case, which we will now focus on, the PBH abundance is exponentially sensitive to the power spectrum amplitude, so there is an additional fine-tuning that we will now quantify. For the simple asymptotic result where we approximate $\sigma^2_{\rm peak} = \mathcal{P}_{\rm peak}$,
\begin{equation}
\beta \sim e^{-\frac{\delta_c^2}{2 \mathcal{P}_\text{peak}}},
\end{equation}
where $\delta_c$ is the formation threshold for PBHs, the fine-tuning on the PBH collapse fraction $\beta$ can be straightforwardly determined analytically using the chain rule,
\begin{equation}
\epsilon_{\beta} = \frac{\delta_c^2}{2 \mathcal{P}_\text{peak}}  \epsilon_{\Ppeak}. \label{eq:Epsilon-beta-RD}
\end{equation}
For the typical value of $\delta_c\simeq0.45$ and a peak amplitude $\mathcal{P}_\text{peak}\sim 10^{-3}$--$10^{-2}$ as required to form PBHs, one has 
\begin{equation}
\frac{\epsilon_\beta}{\epsilon_{\mathcal{P}_{\rm peak}}} = \frac{\delta_c^2}{2\mathcal{P}_\text{peak}} \simeq 10\hspace{0.1em}\text{--}100,
\end{equation}
hence the fine-tuning with respect to the PBH formation rate is 1--2 orders of magnitude worse than the fine-tuning in terms of the peak amplitude of the power spectrum alone.

The full PBH formation calculation is very complicated, and so we do not want to rely on the simple analytical understanding from the asymptotic result above. Therefore, we additionally carry out a detailed numerical study of PBH formation, following the procedure in~\cite{Gow:2021_ACPS}. We include the effects of critical collapse, integrate over all scales in the power spectrum, and additionally include the non-linear relation between $\mathcal{R}$ and $\delta$ \cite{Kawasaki:2019mbl,Young_2019,DeLuca:2019qsy}. For each of the power spectra shown in \cref{sec:Parametric-sensitivity}, we calculate the present-day fraction of dark matter in PBHs, $f_\PBH$, given by
\begin{equation}
f_\PBH = \frac{2}{\Omega_\CDM}\int \diffd(\ln R)\ \frac{R_\text{eq}}{R}\int_{\delta_{R,l}^c}^{\infty} \diffd\delta_{R,l}\ \frac{m}{M_H} P(\delta_{R,l}),
\end{equation}
where $\Omega_\CDM$ is the present-day dark matter density, $R_\text{eq}$ is the horizon scale at matter--radiation equality, $\delta_{R,l}$ is the linear density contrast smoothed on a scale $R$, and the ratio of primordial black hole mass $m$ to horizon mass $M_H$ follows the critical collapse formula,
\begin{equation}
m = KM_H\left(\delta_{R,l} - \frac{3}{8}\delta_{R,l}^2 - \delta_R^c\right)^\gamma,
\end{equation}
where $K = 10$, $\gamma = 0.36$, and the smoothed non-linear density contrast threshold is $\delta_R^c = 0.25$. The PDF $P(\delta_{R,l})$ is Gaussian, with a variance defined in terms of the curvature power spectrum as
\begin{equation}
\sigma^2(R) = \int_{0}^{\infty} \frac{\diffd k}{k}\ \frac{16}{81}(kR)^4 W(kR)\mathcal{P}_\mathcal{R}(k),
\end{equation}
with a window function $W(kR)$, taken as the modified Gaussian window function in~\cite{Gow:2021_ACPS}.

Using this numerical technique, we define another fine-tuning parameter,
\begin{align}\label{eq:pbh_tune}
\epsilon_{f_\PBH} &= \ddv{\log f_\PBH}{\log p},
\intertext{and additionally evaluate the extra fine-tuning to go from the power spectrum amplitude to $f_\PBH$,}
\rho &= \frac{\epsilon_{f_\PBH}}{\epsilon_\mathcal{\Ppeak}}. \label{eq:pbh_rho}
\end{align}
These quantities are displayed in \cref{tab:Mishra-fine-tuning,tab:Hertzberg-fine-tuning,tab:Cicoli-fine-tuning,tab:Germani-fine-tuning}, and show that the analytical calculation is robust in the sense that the ratio $\rho$ varies between 22 and 38 in the 4 models (and parameter choices) we consider.

The fact that this is not ``very'' fine-tuned, \textit{e.g.}~compared to the $10^4$--$10^5$ tuning we find for the sensitivity of the peak of the power spectrum to variations of the parameters for the potential \cref{eq:Mishra-potential}, shows that (perhaps surprisingly) the main reason why PBH production is so fine-tuned is that the power spectrum amplitude is so sensitive to the duration of USR (and \textit{e.g.}~the width of the flat part of the potential around the feature), rather than the fact that the PBH production is exponentially sensitive to the amplitude. For example, the early universe might not have been radiation dominated \cite{1981SvA....25..406P,Harada:2016mhb,Carr:2017edp,Cole:2017gle,Dalianis:2018ymb,Allahverdi:2020bys,Dienes:2021woi,Bhattacharya:2023ztw,Ballesteros_2020} and in an early matter dominated era the PBH fraction may change to \cite{Harada:2016mhb}
\begin{equation}
\beta \propto \mathcal{P}_\text{peak}^{5/4}
\end{equation}
with 
\begin{equation}
\ddv{\ln\beta}{\ln \mathcal{P}_\text{peak}} = \frac{5}{4}.
\end{equation}
This is 1--2 orders of magnitude less fine-tuned than the value in a radiation era (and independent of the power spectrum amplitude), so the total fine-tuning of $\beta$ based on the superposed feature potential in \cref{eq:Mishra-potential} will be $10^4$--$10^5$ in matter domination vs $10^5$--$10^7$ in radiation domination.

 A caveat to both the analytical and numerical calculations here is that they assume the curvature perturbations follow a Gaussian distribution. It has been demonstrated that non-Gaussianity, typically treated perturbatively, can have a significant impact on PBH formation~\cite{Bullock:1996_Non-Gaussian,Young:2015_Influence,Yoo:2019_Abundance,Palma:2020_Non-Gaussian,Taoso:2021_Non-gaussianities,Young:2022_non-G}. One source of non-Gaussianity that has gained recent interest is the possible presence of quantum diffusion in USR inflationary models, which can be handled using the stochastic formalism, typically resulting in non-Gaussian tails which must be treated non-perturbatively~\cite{Fujita:2013_Algorithm,Vennin:2015_Correlation,Pattison:2017_Quantum,Ezquiaga:2019_Exponential,Figueroa:2020_Non-Gaussian,Ando:2020_Power,Pattison:2021_USR,Rigopoulos:2021_Inflation,Tada:2021_Statistics,Jackson:2022_Numerical,Gow:2022jfb,Ferrante:2022mui}. The presence of primordial non-Gaussianity alters the amount of fine-tuning between the power spectrum amplitude and $f_\PBH$, with $f_\PBH\propto e^{\left(\delta_c/(\sqrt{2}\sigma)\right)^p}$~\cite{Nakama:2016kfq,Nakama:2019htb} where $p=2$ in the Gaussian case. However, even in the extreme (non-perturbative) limit of $\chi$-squared perturbations ($p=1$)  the fine-tuning would then only be reduced by a square root and factor of two, so we conclude that even large non-Gaussianity is unlikely to significantly ameliorate this fine-tuning. There is also the possibility that non-Gaussianity can reduce the power spectrum amplitude required to generate a significant abundance of PBHs, which could reduce the fine tuning. However, the only way to properly determine the fine-tuning for PBH formation in models with quantum diffusion is to carry out a full calculation, which is beyond the scope of this work. We also caution that local non-Gaussianity tends to generate unacceptably large isocurvature perturbations unless $|f_\PBH|\ll 1$~\cite{Tada:2015noa,Young:2014oea,vanLaak:2023ppj}.

\section{Audit summary and concluding remarks}
\label{sec:Conclusions}

As we have quantified in the preceding sections, each class of single-field models that we have examined require a high degree of parametric tuning to generate a significant number of PBHs within a certain mass range whilst simultaneously satisfying large scale observational constraints. The precise degree of tuning varies across the classes considered (the difference between some classes being artificial to some extent \textit{cf}.~\cref{fn:Canonical-kinetic}), with the polynomial class discussed in \cref{tab:Hertzberg-fine-tuning} requiring an additional tuning of initial conditions.  On top of this, the fine-tuning of an infinite number of coefficients of the potential beyond quintic must also be considered since the field range of interest in this example is such that $\Delta\phi/\Lambda > 1$, where $\Lambda$ would ordinarily suppress higher-dimensional operators at small-field values. Order one changes in the peak amplitude of the primordial power spectrum require the precision of potential parameters to range from one part in a few hundred to one part in a hundred million depending on the potential class considered, as detailed in the previous section. However, it is interesting to note that the level of fine-tuning is generally comparable amongst parameters for a chosen model.

Given that the PBH abundance is exponentially sensitive to the amplitude of the primordial power spectrum, it is no surprise that some fine-tuning is required to generate an interesting abundance of PBHs (\textit{i.e.}~not fewer than one per Hubble volume today~\cite{Cole:2017gle}, and not more than the observed dark matter density). This conclusion is well known and was studied in detail in~\cite{Nakama:2018utx}. However, we quantify in this work that the fine-tuning is more severe than concluded by~\cite{Nakama:2018utx}, in part because the peak amplitude of the power spectrum is itself exponentially sensitive to the duration of ultra-slow-roll inflation (see \textit{e.g.}~\cite{Hertzberg:2017dkh,Passaglia_2019}). This implies that the PBH abundance is (at least) double-exponentially sensitive to the inflationary parameters which determine the existence and duration of the USR phase. The PBH abundance at any given scale is sensitive by two more orders of magnitude (with respect to the power spectrum amplitude) to the parameters of the potentials assuming that the PBHs form in radiation-domination. The larger contribution to the fine-tuning is from the sensitivity of the power spectrum amplitude to the potential parameters, as opposed to from the power spectrum to the PBH abundance. This means that generating an amplitude of secondary stochastic gravitational waves~\cite{Domenech:2021ztg} that might be observable with, for example, pulsar timing arrays or future gravitational wave detectors such as LISA, requires significant fine-tuning.

Whether one views this tuning as acceptable or not is a matter of taste to some extent, itself a manifestation of unspoken priors (\textit{cf}.~the discussion in \cref{app:Fine-tuning-criteria}). However, there are two criteria that all of these potential classes must necessarily satisfy: whether they can be realised at the level of the effective action and also respect tensor constraints on large scales. Of the three classes of potentials that we study, deceleration via feature overshoot suffers from the feature being artificially added `by hand' and is problematic from the perspective of being realisable at the level of the effective potential. Polynomial potentials are generally considered realisable as effective potentials, but they produce tensor-to-scalar ratios that are disallowed by large-scale CMB measurements, unless one sacrifices the desire for the model to be insensitive to its initial field value, in which case this can be avoided by allowing inflation to begin off-attractor. Finally non-polynomial feature potentials can be motivated by high energy theories, but again struggle to obey CMB constraints on large scales.

Furthermore, the fact that the background effective potential, which is the zero mode of the effective action, is necessarily convex~\cite{Peskin:1995ev, Haymaker:1983xk, Amer:1982ri} should give caution to designing any classical function to produce the desired effect without accounting for additional degrees of freedom that can allow for the desired non-convexity at the relevant scales. None of the potentials considered are convex.
Even before adding a feature to generate PBHs, CMB observations of a red spectral index and small tensor-to-scalar ratio favour a (concave) potential with $V''<0$. This is also in tension with a potential which expands to look like a monomial far away from the feature, which leads to the many difficulties of designing a polynomial potential. 
In general, producing light PBHs is likely to be easier because they exit on scales far removed from those which generate the CMB and hence the addition of a feature such as an inflection point on such scales is less likely to ruin the predictions on CMB scales. This was discussed in \textit{e.g.}~\cite{Ballesteros:2017fsr,Ballesteros:2020sre}. Of the potentials that we examine in this paper, only the Mishra and Sahni potential (with our choice of fiducial parameter sets, see \cref{tab:Mishra-fiducial,tab:Germani-fiducial}) are in good agreement with observational data of the spectral index at the CMB pivot scale. We present in \cref{tab:ns} both the spectral index $n_s=-2\epsilon_H-\eta_H+1$ and tensor-to-scalar ratio $r=16\epsilon_H$, calculated at $\phi_{\rm CMB}$ as defined in the main text for each potential. For reference, the Planck constraint on the spectral index is $n_s=0.9649\pm0.0042$ at the 68\% confidence level~\cite{Planck:2018jri} and the bound on the tensor-to-scalar ratio including BICEP--Keck data is $r < 0.032$ at the 95\% confidence level~\cite{Tristram:2021tvh}. All four potentials satisfy the tensor-to-scalar ratio bound.
\begin{table}[H]
\caption{Values of spectral index and tensor-to-scalar ratio}
\label{tab:ns}
\centering
\begin{tabular}{ccccc}
	 &Mishra and Sahni & Hertzberg and Yamada & Germani and Prokopec & Cicoli \textit{et al}. \\ \hline
	  $n_s$& 0.9648 & 0.9820 & 0.9567 & 0.9400  \\
	  $r$& 0.0026 & $4.8\times10^{-7} $& 0.0063 & 0.018 \\ \hline
\end{tabular}
\end{table}

We therefore conclude that in contrast to the ``WIMP miracle'' there is instead a significant challenge to explain why the PBH abundance is not either zero or exponentially too large. Even the addition of perturbative non-Gaussianity would not significantly change this conclusion, since the PBH abundance is still highly sensitive to the power spectrum amplitude. However, our conclusions are specific to the formation of PBHs generated by the direct collapse of large amplitude perturbations shortly after horizon entry following a period of single-field inflation, and it would be interesting to determine whether alternative inflationary scenarios  \textit{e.g.}~\cite{Yokoyama:1995ex,Cai_2018,KAMENSHCHIK2019201,Ashoorioon:2020hln,romano2020sound,Palma:2020_Seeding,Kawai:2021edk,Fumagalli:2020_Turning, Geller:2022nkr, Iacconi_2022,Talebian:2022cwk,Papanikolaou:2022did, Ozsoy:2023ryl} and/or alternative PBH formation scenarios require less fine-tuning \textit{e.g.}~\cite{Martin:2019nuw,Dvali:2021byy,Animali_2023}.

\section*{Acknowledgements}
The authors thank Guillermo Ballesteros, Matteo Braglia, Swagat Mishra and the authors of \cite{Qin:2023lgo}. CB and SP thank the organisers of the Messengers of the Early Universe: Gravitational Waves and Primordial Black Holes workshop in Padua where an early version of this work was first presented. PC acknowledges support from the Institute of Physics at the University of Amsterdam. AG acknowledges support from the Science and Technology Facilities Council [grant numbers ST/S000550/1 and ST/W001225/1]. CB acknowledges support from the Science and Technology Facilities Council [grant number ST/T000473/1]. For the purpose of open access, the authors have applied a Creative Commons Attribution (CC BY) licence to any Author Accepted Manuscript version arising. No new raw data were generated or analysed in support of this research.

\appendix
\renewcommand\thefigure{A.\arabic{figure}}
\renewcommand{\theHfigure}{A.\arabic{figure}}
\setcounter{figure}{0}

\section{Fine-tuning criteria}
\label{app:Fine-tuning-criteria}

It is not possible to decide upon a measure or a set of criteria for fine-tuning that is free from ambiguities. Operationally, the question one is trying to determine is the degree of sensitivity of certain observables to small changes in the parameters of a given model construction, typically specified by parameters (Wilson coefficients) of some Lagrangian, itself supposed to be understood as a mere bootstrap to the full quantum effective action. The problem here already, is that Wilson coefficients by themselves do not correspond to physical observables. In the context of particle physics where one can freely presume the existence of an S-matrix, it is well understood that only on shell S-matrix elements are observable, and Lagrangians are only a calculational means to obtain them\footnote{This is the content of the so-called equivalence theorem~\cite{Coleman:1969sm}. Simply put, one is free to make arbitrary (non-singular) field redefinitions to a given Lagrangian and still end up with the same on-shell S-matrix. The canonical textbook example is to take a free scalar field theory $\mathcal{L} = -\frac{1}{2}(\partial\phi)^2 - \frac{m^2}{2}\phi^2$ and make an arbitrary field redefinition $\phi = f(\psi)$ so that $\mathcal{L} = -\frac{1}{2}f'(\psi)^2(\partial\psi)^2 - \frac{m^2}{2}f(\psi)^2$. For an arbitrary $f(\psi)$, one potentially ends up with a complicated interacting Lagrangian. Of course, all diagrams entering any given scattering process sum to zero as they must for a free theory, which might seem like a remarkable series of cancellations if one didn't know better.}. Specifically, the coefficients of a Lagrangian can be freely redefined and mix into each other under field redefinitions whilst leaving observable quantities invariant. In the context of coupling fields to gravity, one might seemingly have the added subtlety of needing to specify the \textit{frame} in which one considers certain Wilson coefficients to be `naturally' order unity. What is an order unity Wilson coefficient in the Jordan frame, for instance, becomes exponentially suppressed in the Einstein frame for non-minimal couplings of the form $\xi\phi^2 R$. However when it comes to estimating the degree of tuning of a parameter, standard power counting arguments direct one to work in the Einstein frame\footnote{The reason for this is needing to work with canonically normalised fields when doing standard power counting in effective field theory~\cite{Burgess:2007pt}. One is of course free to not normalise fields, but then the non-canonical nature of the kinetic terms modifies the power counting in a complicated manner such that the final conclusions expressed in terms of observable quantities will remain unchanged if everything is kept track of properly.}. 

Even with this series of caveats in mind, one is still left with the intractable issue of the underlying \textit{measure problem} when trying to determine how `tuned' a given parameter is. Specifically, a flat prior for a given Wilson coefficient translates into a logarithmic prior had we chosen to parameterise it as the exponential of some other parameter. Moreover, the choice of a flat prior itself is presuming something about the ultra-violet completion of the low energy theory. For instance, whether neutrino masses (determined by the coefficient of the dimension five Weinberg operator in the Standard Model effective theory~\cite{Brivio:2017vri}) are to be assigned a flat logarithmic prior, or some other prior when doing cosmological inference with necessarily limited data can lead to conflicting conclusions for the hierarchy of neutrino masses~\cite{Simpson:2017qvj,Schwetz:2017fey}.
   
This somewhat unsatisfactory state of affairs is of direct relevance to quantifying the fine-tuning of the abundance and mass function of PBHs. Nevertheless, one is free to stick to one choice for the restricted purpose of comparing between different model constructions, which is the perspective we adopt in the paper. We comment here on an alternative fine-tuning criterion in the literature, and discuss how this relates to the measures we primarily use in \cref{sec:Fine-tuning}. Nakama and Wang~\cite{Nakama:2018utx} introduced the fine-tuning measure 
\begin{equation}
\epsilon_{\rm NW} = \frac{\xmax-\xmin}{(\xmax+\xmin)/2} \label{eq:Epsilon-NW}
\end{equation}
which they apply to the amplitude of the primordial perturbations $x=\sigma\sim \sqrt{\calP_\calR}$ and the minimum and maximum values are the limits required to generate a certain range of $f_\PBH$. Similarly to \cref{eq:Fine-tuning-parameter}, this definition takes no account of the possible range of values which the parameter $x$ could take in principle. Defining the fine-tuning via \cref{eq:Epsilon-NW} is equivalent to choosing a uniform prior from 0 to $\xmax>0$ and hence the fine-tuning to reach a power spectrum amplitude between $A/2$ and $A$ is independent of $A$, which makes no reference to the observed amplitude on CMB scales. This means equal fine-tuning values are assigned to the power spectrum being in the range $0.5\times10^{-9}$--$10^{-9}$ or \textit{e.g.}~$0.5\times10^{-3}$--$10^{-3}$ on some arbitrary small scale, despite only the latter range requiring a special feature in the potential.

We instead use the fine-tuning definition of Azhar and Loeb~\cite{Azhar:2018lzd} (see also~\cite{Franciolini:2018vbk} who use the same definition in a related context) who considered fine-tuning in `evading' the $\fpbh$ constraints by fitting a lognormal mass function with two free parameters (the central mass and width) to a set of constraints derived assuming a monochromatic mass spectrum. 
If there is more than one important model parameter then the definition \cref{eq:Fine-tuning-parameter} could be extended to add the derivative of all parameters in quadrature. We also comment that whilst we have typically found the fine-tuning amplitude to have the same order of magnitude between all relevant parameters, that there are expected to be degeneracy directions in which a particular combination of parameters leads to the same duration of USR and hence a comparable peak height.

This definition does not depend on priors, but this does not imply that our interpretation of the results should be prior independent. If you had a good reason to believe that a parameter is tightly constrained then a large value of $\epsilon$ might not be concerning, but in practise we normally do not have a theoretical motivation to take a very narrow prior.

\section{Analytic determination of the fine-tuning}

In general the power spectrum has to be calculated numerically in models of inflation which break slow roll. However, analytic formulae exist for the idealised case of a completely flat potential, which generates ``pure'' ultra-slow-roll (USR) inflation, with
\begin{equation}
\dot{\phi}(N)=\dot{\phi}_i e^{-3N}.
\end{equation}
From the equation of motion above one can derive the following relation between the width of the flat section $\Delta\phi$ and the duration $N_\text{USR}$ (measured in e-folds),
\begin{equation}
\Delta\phi = \frac{\dot{\phi_i}}{3H}\left(1-e^{-3\Nusr}\right)=\frac{\sqrt{2\epsilon_{H,i}}}{3}\left(1-e^{-3\Nusr}\right), \label{eq:Delta-phi}
\end{equation}
where $\dot{\phi}$, or $\epsilon_{H}\equiv-\dot{H}/H^2=\dot{\phi}^2/(2 H^2 )$, should be evaluated as USR begins (denoted with a subscript $i$). There is a maximum distance the inflaton can ever (classically) roll, which is
\begin{equation}
\Delta\phi_{\rm max}\equiv \int_0^\infty \ddv{\phi}{N} \diffd N =\frac{\dot{\phi}_i}{H}\int_0^\infty e^{-3N} \diffd N = \frac{\dot{\phi}_i}{3H} = \frac{\sqrt{2\epsilon_{H,i}}}{3} \, ,
\end{equation}
and we introduce the convenient small (positive) parameter which measures how close the inflaton comes to rolling this maximum distance
\begin{equation}
f\equiv\frac{\Delta\phi_{\rm max}-\Delta\phi}{\Delta\phi_{\rm max}}\ll1.
\end{equation}
The increase in the power spectrum can be approximately determined using
\begin{equation}
\frac{\Ppeak}{A_s} \simeq \frac{\epsilon_{H,i}}{\epsilon_{H,f}} \simeq e^{6\Nusr},
\end{equation}
where $A_s\simeq2\times10^{-9}$ is the amplitude of the power spectrum on CMB scales and $\epsilon_{H,f}$ is evaluated at the end of USR, which in this model is the minimum value of $\epsilon_H$. We can invert \cref{eq:Delta-phi} to find the peak power spectrum amplitude is
\begin{equation}
\Ppeak \simeq A_s \left(1 - \frac{3}{\sqrt{2\epsilon_{H,i}}}\Delta\phi\right)^{-2}= A_s \frac{1}{f^2}. \label{eq:P-peak}
\end{equation}

The fine-tuning of the peak amplitude in the high peak limit ($f\ll1$) is given by
\begin{equation}
\epsilon_{\Ppeak}=\frac2f \frac{\Delta\phi}{\Delta\phi_{\rm max}} \simeq \frac2f, \end{equation}
which diverges in the limit $f\rightarrow0$ corresponding to an infinitely high peak. Using \cref{eq:P-peak} one can determine the simple relation between the fine-tuning and peak height
\begin{equation}
\epsilon_{\Ppeak} \simeq 2 \sqrt{\frac{\Ppeak}{A_s}}\simeq 3\times10^3  \sqrt{\frac{\Ppeak}{5\times 10^{-3}}}.
\end{equation}
Hence we see that larger peaks require more fine-tuning, with the fine-tuning being proportional to the square root of the peak amplitude. We have empirically found this relation to be approximately true for all of the four potentials we studied numerically, with corrections typically of order $10\%$ when varying the peak amplitude by two orders of magnitude. For the peak height we have studied of $5\times10^{-3}$, relevant for PBH formation, the fine-tuning value is $\epsilon_{\Ppeak}\simeq3\times10^3$, which is towards the lower end of the range of values found for the smooth potentials we have tested. 
We note that the fine-tuning amplitude is independent of the position of the flat feature in the potential, which means it is independent of the corresponding PBH mass.

We now consider a potential proposed by Starobinsky~\cite{Starobinsky:1992ts} which is continuous but has a discontinuous first derivative, being $V_1'$ before the step and $V_2'\equiv V_1'/\alpha$ afterwards, with the steepness after the step being much less (corresponding to $\alpha\gg1$) and hence there being a period of USR until the inflaton has lost sufficient kinetic energy to reach the SR attractor value of $\dot{\phi}_2=-V_2'/(3H)$. Before the step $\dot{\phi}_1=-V_1'/(3H)$ and hence the duration of USR is
\begin{equation}
\Nusr = \frac13 \ln\left(\frac{V_1'}{V_2'}\right)=\frac13\ln(\alpha),
\end{equation} 
and this will lead to an associated boost of the power spectrum (albeit a plateau rather than peak) by
\begin{equation}
e^{6\Nusr} =\left(\frac{V_1'}{V_2'}\right)^2=\frac{\epsilon_1}{\epsilon_2}=\alpha^2.
\end{equation}
The fine-tuning parameter is hence
\begin{equation}
\epsilon_{\Ppeak}=2
\end{equation}
which is extremely small, and (uniquely amongst the models we considered) is independent of the amplitude of the peak, at least in the high peak limit corresponding to $\alpha\gg1$.

We note that an analytic formula for the power spectrum exists~\cite{Pi:2022zxs} and using this result would not change our conclusions. However, we comment that this model does not work without modification since the power spectrum does not decrease again on small scales and the instant transition in the derivative, apart from being unrealistic, also leads to large oscillations. This model is also unique in having a zero second derivative of the potential which is connected to the lack of a constant roll period after USR ends and the fact that the power spectrum has more of a plateau than a peak.

\section[\texorpdfstring{Power spectrum features from rapid changes in $H$}{Power spectrum features from rapid changes in H}]{Power spectrum features from rapid changes in $\bm{H}$}
\begin{figure}[H]
\centering
\includegraphics[width=0.5\textwidth]{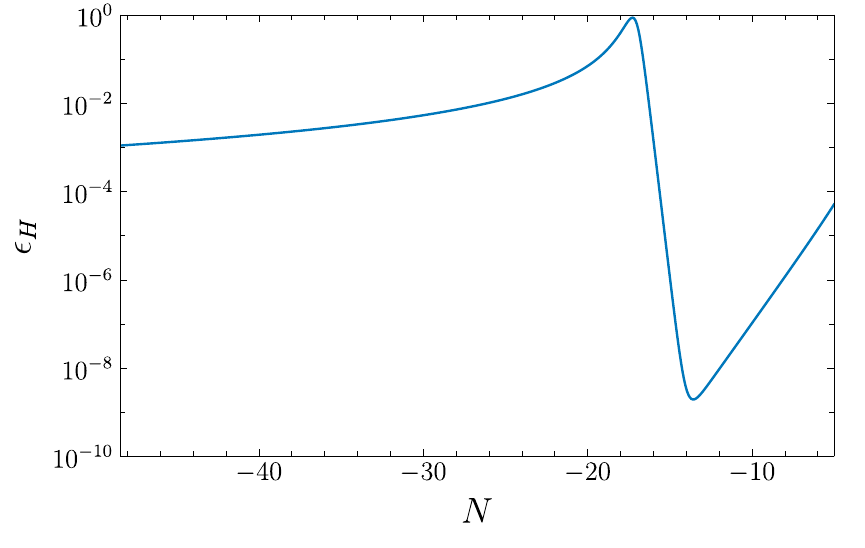}\includegraphics[width=0.5\textwidth]{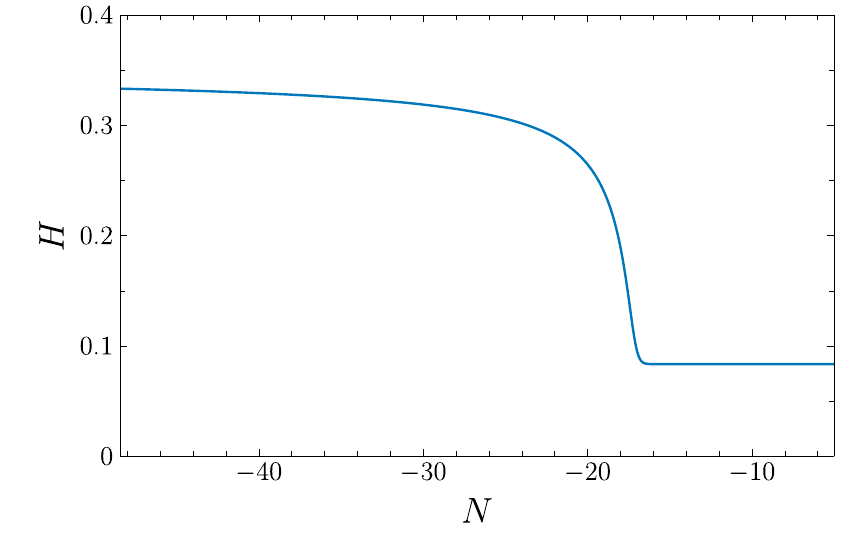}
\includegraphics[width=0.5\textwidth]{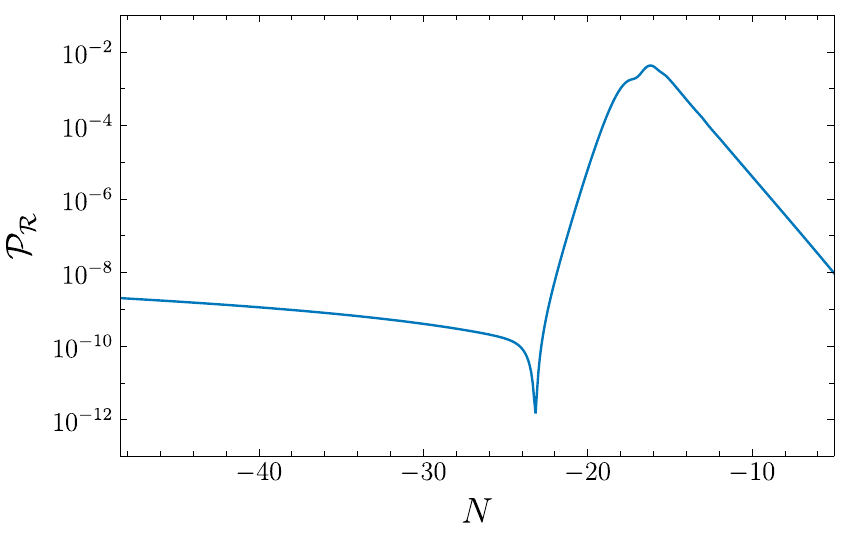}\includegraphics[width=0.5\textwidth]{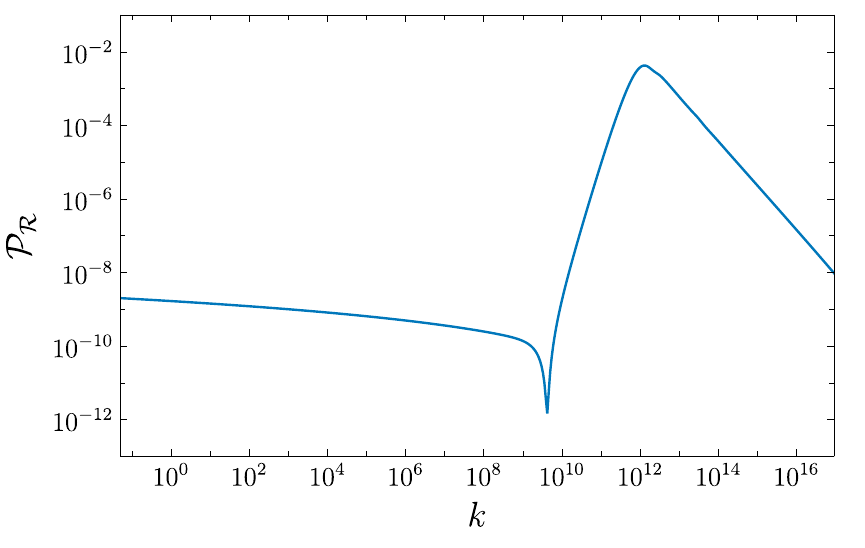}
\caption{\textit{Bottom left:} Demonstration of a bump in the peak of the power spectrum when plotted over $N$, due to a rapid drop in $H$ (or equivalently $\epsilon_H \simeq 1$) where the $N$ range is normalised to the end of inflation. \textit{Bottom right:} The feature disappears when the observable $k$ is used as the plotting variable. \textit{Top left:} The corresponding first slow-roll parameter plotted as a function of $N$. \textit{Top right:} The corresponding value of $H$ plotted as a function of $N$.}
\label{fig:Cicoli-feature}
\end{figure}

When plotting the power spectrum, it is common to use e-folds $N$ rather than $k$, to connect back to the inflationary evolution. However, in certain cases this can cause spurious features to appear in the power spectrum plot that may mislead readers. One example of this is the model discussed in~\cite{Cicoli:2018asa}, \cref{eq:Cicoli-potential}. While it is typical for $\epsilon_H$ to grow before the USR phase, in this model it gets very close to one, corresponding to a rapid drop in $H$ (see \cref{fig:Cicoli-feature}, top row). This results in a kink to the left of the peak of the power spectrum, as can be seen in the bottom left panel of \cref{fig:Cicoli-feature}, which could be seen as a motivation for features in observables such as the PBH mass distribution and scalar-induced stochastic gravitational-wave background. However, when transforming to $k = a(N)H(N) \neq \exp(N)$, there is a corresponding kink in the $k(N)$ relation, meaning that the feature disappears from the power spectrum when plotted over $k$. It should also be noted that the kink appears near the peak of the power spectrum, despite the corresponding $\epsilon\simeq1$  phase appearing before the onset of USR.

\bibliographystyle{JHEP-edit} 
\bibliography{PBH_model_building_paper_2}{}

\providecommand{\href}[2]{#2}\begingroup\raggedright\begin{thebibliography}{100}

\bibitem{Zeldovich:1969ff}
{\relax Ya}.~B. Zeldovich and R.~A. Sunyaev, \emph{The interaction of matter
  and radiation in a hot-model universe},
  \href{https://doi.org/10.1007/BF00661821}{\emph{Astrophysics and Space
  Science} {\bfseries 4}{\bfseries (3)} (1969) 301}.

\bibitem{Garcia_Bellido_1996}
J.~García-Bellido, A.~Linde and D.~Wands, \emph{Density perturbations and
  black hole formation in hybrid inflation},
  \href{https://doi.org/10.1103/physrevd.54.6040}{\emph{Physical Review D}
  {\bfseries 54}{\bfseries (10)} (1996) 6040}
  [\href{https://arxiv.org/abs/astro-ph/9605094}{{\ttfamily
  astro-ph/9605094}}].

\bibitem{Yokoyama:1995ex}
J.~Yokoyama, \emph{{Formation of MACHO primordial black holes in inflationary
  cosmology}},
  \href{https://ui.adsabs.harvard.edu/abs/1997A%26A...318..673Y/abstract}{\emph{Astronomy
  \& Astrophysics} {\bfseries 318} (1997) 673}
  [\href{https://arxiv.org/abs/astro-ph/9509027}{{\ttfamily
  astro-ph/9509027}}].

\bibitem{Ivanov_1998}
P.~Ivanov, \emph{Nonlinear metric perturbations and production of primordial
  black holes}, \href{https://doi.org/10.1103/physrevd.57.7145}{\emph{Physical
  Review D} {\bfseries 57}{\bfseries (12)} (1998) 7145}
  [\href{https://arxiv.org/abs/astro-ph/9708224}{{\ttfamily
  astro-ph/9708224}}].

\bibitem{Green:2020_Primordial}
A.~M. Green and B.~J. Kavanagh, \emph{Primordial black holes as a dark matter
  candidate}, \href{https://doi.org/10.1088/1361-6471/abc534}{\emph{Journal of
  Physics G} {\bfseries 48}{\bfseries (4)} (2021) 043001}
  [\href{https://arxiv.org/abs/2007.10722}{{\ttfamily 2007.10722}}].

\bibitem{Escriva:2022duf}
A.~Escrivà, F.~Kuhnel and Y.~Tada, \emph{{Primordial Black Holes}},  (2022)
  [\href{https://arxiv.org/abs/2211.05767}{{\ttfamily 2211.05767}}].

\bibitem{Buttazzo:2013uya}
D.~Buttazzo \textit{et~al}., \emph{{Investigating the near-criticality of the
  Higgs boson}}, \href{https://doi.org/10.1007/JHEP12(2013)089}{\emph{Journal
  of High Energy Physics} {\bfseries 2013}{\bfseries (12)} (2013) 089}
  [\href{https://arxiv.org/abs/1307.3536}{{\ttfamily 1307.3536}}].

\bibitem{Baumann:2009ds}
D.~Baumann, \emph{{The Eta Problem}},  in \emph{{TASI Lectures on Inflation}},
  p.~103, 2011, \href{https://arxiv.org/abs/0907.5424}{{\ttfamily 0907.5424}}.

\bibitem{Byrnes:2019_Steepest}
C.~T. Byrnes, P.~S. Cole and S.~P. Patil, \emph{Steepest growth of the power
  spectrum and primordial black holes},
  \href{https://doi.org/10.1088/1475-7516/2019/06/028}{\emph{Journal of
  Cosmology and Astroparticle Physics} {\bfseries 2019}{\bfseries (06)} (2019)
  028} [\href{https://arxiv.org/abs/1811.11158}{{\ttfamily 1811.11158}}].

\bibitem{Ivanov:1994pa}
P.~Ivanov, P.~Naselsky and I.~Novikov, \emph{Inflation and primordial black
  holes as dark matter},
  \href{https://doi.org/10.1103/PhysRevD.50.7173}{\emph{Physical Review D}
  {\bfseries 50}{\bfseries (12)} (1994) 7173}.

\bibitem{Kinney:2005vj}
W.~H. Kinney, \emph{Horizon crossing and inflation with large $\eta$},
  \href{https://doi.org/10.1103/PhysRevD.72.023515}{\emph{Physical Review D}
  {\bfseries 72}{\bfseries (2)} (2005) 023515}
  [\href{https://arxiv.org/abs/gr-qc/0503017}{{\ttfamily gr-qc/0503017}}].

\bibitem{Garcia-Bellido:2017mdw}
J.~García-Bellido and E.~Ruiz~Morales, \emph{Primordial black holes from
  single field models of inflation},
  \href{https://doi.org/10.1016/j.dark.2017.09.007}{\emph{Physics of the Dark
  Universe} {\bfseries 18} (2017) 47}
  [\href{https://arxiv.org/abs/1702.03901}{{\ttfamily 1702.03901}}].

\bibitem{Dimopoulos:2017ged}
K.~Dimopoulos, \emph{Ultra slow-roll inflation demystified},
  \href{https://doi.org/10.1016/j.physletb.2017.10.066}{\emph{Physics Letters
  B} {\bfseries 775} (2017) 262}
  [\href{https://arxiv.org/abs/1707.05644}{{\ttfamily 1707.05644}}].

\bibitem{Motohashi:2017kbs}
H.~Motohashi and W.~Hu, \emph{Primordial black holes and slow-roll violation},
  \href{https://doi.org/10.1103/PhysRevD.96.063503}{\emph{Physical Review D}
  {\bfseries 96}{\bfseries (6)} (2017) 063503}
  [\href{https://arxiv.org/abs/1706.06784}{{\ttfamily 1706.06784}}].

\bibitem{Kannike:2017bxn}
K.~Kannike, L.~Marzola, M.~Raidal and H.~Veermäe, \emph{Single field double
  inflation and primordial black holes},
  \href{https://doi.org/10.1088/1475-7516/2017/09/020}{\emph{Journal of
  Cosmology and Astroparticle Physics} {\bfseries 2017}{\bfseries (09)} (2017)
  020} [\href{https://arxiv.org/abs/1705.06225}{{\ttfamily 1705.06225}}].

\bibitem{Liu:2020oqe}
J.~Liu, Z.-K. Guo and R.-G. Cai, \emph{Analytical approximation of the scalar
  spectrum in the ultraslow-roll inflationary models},
  \href{https://doi.org/10.1103/PhysRevD.101.083535}{\emph{Physical Review D}
  {\bfseries 101}{\bfseries (8)} (2020) 083535}
  [\href{https://arxiv.org/abs/2003.02075}{{\ttfamily 2003.02075}}].

\bibitem{Karam:2022nym}
A.~Karam \textit{et~al}., \emph{Anatomy of single-field inflationary models for
  primordial black holes},
  \href{https://doi.org/10.1088/1475-7516/2023/03/013}{\emph{Journal of
  Cosmology and Astroparticle Physics} {\bfseries 2023}{\bfseries (03)} (2023)
  013} [\href{https://arxiv.org/abs/2205.13540}{{\ttfamily 2205.13540}}].

\bibitem{Cheung:2007st}
C.~Cheung \textit{et~al}., \emph{The effective field theory of inflation},
  \href{https://doi.org/10.1088/1126-6708/2008/03/014}{\emph{Journal of High
  Energy Physics} {\bfseries 2008}{\bfseries (03)} (2008) 014}
  [\href{https://arxiv.org/abs/0709.0293}{{\ttfamily 0709.0293}}].

\bibitem{Senatore:2010wk}
L.~Senatore and M.~Zaldarriaga, \emph{The effective field theory of multifield
  inflation}, \href{https://doi.org/10.1007/JHEP04(2012)024}{\emph{Journal of
  High Energy Physics} {\bfseries 2012}{\bfseries (04)} (2012) 024}
  [\href{https://arxiv.org/abs/1009.2093}{{\ttfamily 1009.2093}}].

\bibitem{Achucarro:2012sm}
A.~Achúcarro \textit{et~al}., \emph{Effective theories of single field
  inflation when heavy fields matter},
  \href{https://doi.org/10.1007/JHEP05(2012)066}{\emph{Journal of High Energy
  Physics} {\bfseries 2012}{\bfseries (05)} (2012) 066}
  [\href{https://arxiv.org/abs/1201.6342}{{\ttfamily 1201.6342}}].

\bibitem{Chluba:2015bqa}
J.~Chluba, J.~Hamann and S.~P. Patil, \emph{{Features and new physical scales
  in primordial observables: Theory and observation}},
  \href{https://doi.org/10.1142/S0218271815300232}{\emph{International Journal
  of Modern Physics D} {\bfseries 24}{\bfseries (10)} (2015) 1530023}
  [\href{https://arxiv.org/abs/1505.01834}{{\ttfamily 1505.01834}}].

\bibitem{Cole:2022xqc}
P.~S. Cole, A.~D. Gow, C.~T. Byrnes and S.~P. Patil, \emph{Steepest growth
  re-examined: repercussions for primordial black hole formation},  (2022)
  [\href{https://arxiv.org/abs/2204.07573}{{\ttfamily 2204.07573}}].

\bibitem{Felder:1998vq}
G.~Felder, L.~Kofman and A.~Linde, \emph{Instant preheating},
  \href{https://doi.org/10.1103/PhysRevD.59.123523}{\emph{Physical Review D}
  {\bfseries 59}{\bfseries (12)} (1999) 123523}
  [\href{https://arxiv.org/abs/hep-ph/9812289}{{\ttfamily hep-ph/9812289}}].

\bibitem{Sasaki:2018_Primordial}
M.~Sasaki, T.~Suyama, T.~Tanaka and S.~Yokoyama, \emph{Primordial black
  holes---perspectives in gravitational wave astronomy},
  \href{https://doi.org/10.1088/1361-6382/aaa7b4}{\emph{Classical and Quantum
  Gravity} {\bfseries 35}{\bfseries (6)} (2018) 063001}
  [\href{https://arxiv.org/abs/1801.05235}{{\ttfamily 1801.05235}}].

\bibitem{Carr:2020xqk}
B.~Carr and F.~Kühnel, \emph{{Primordial Black Holes as Dark Matter: Recent
  Developments}},
  \href{https://doi.org/10.1146/annurev-nucl-050520-125911}{\emph{Annual Review
  of Nuclear and Particle Science} {\bfseries 70} (2020) 355}
  [\href{https://arxiv.org/abs/2006.02838}{{\ttfamily 2006.02838}}].

\bibitem{Bhattacharya:2023ztw}
S.~Bhattacharya, \emph{{Primordial Black Hole Formation in Non-Standard
  Post-Inflationary Epochs}},
  \href{https://doi.org/10.3390/galaxies11010035}{\emph{Galaxies} {\bfseries
  11}{\bfseries (1)} (2023) 35}
  [\href{https://arxiv.org/abs/2302.12690}{{\ttfamily 2302.12690}}].

\bibitem{Azhar:2018lzd}
F.~Azhar and A.~Loeb, \emph{Gauging fine-tuning},
  \href{https://doi.org/10.1103/PhysRevD.98.103018}{\emph{Physical Review D}
  {\bfseries 98}{\bfseries (10)} (2018) 103018}
  [\href{https://arxiv.org/abs/1809.06220}{{\ttfamily 1809.06220}}].

\bibitem{Hertzberg:2017dkh}
M.~P. Hertzberg and M.~Yamada, \emph{Primordial black holes from polynomial
  potentials in single field inflation},
  \href{https://doi.org/10.1103/PhysRevD.97.083509}{\emph{Physical Review D}
  {\bfseries 97}{\bfseries (8)} (2018) 083509}
  [\href{https://arxiv.org/abs/1712.09750}{{\ttfamily 1712.09750}}].

\bibitem{Nakama:2018utx}
T.~Nakama and Y.~Wang, \emph{Do we need fine-tuning to create primordial black
  holes?}, \href{https://doi.org/10.1103/PhysRevD.99.023504}{\emph{Physical
  Review D} {\bfseries 99}{\bfseries (2)} (2019) 023504}
  [\href{https://arxiv.org/abs/1811.01126}{{\ttfamily 1811.01126}}].

\bibitem{Carr:2019hud}
B.~Carr, S.~Clesse and J.~García-Bellido, \emph{{Primordial black holes from
  the QCD epoch: linking dark matter, baryogenesis and anthropic selection}},
  \href{https://doi.org/10.1093/mnras/staa3726}{\emph{Monthly Notices of the
  Royal Astronomical Society} {\bfseries 501}{\bfseries (1)} (2021) 1426}
  [\href{https://arxiv.org/abs/1904.02129}{{\ttfamily 1904.02129}}].

\bibitem{Braglia:2022phb}
M.~Braglia, A.~Linde, R.~Kallosh and F.~Finelli, \emph{Hybrid
  $\alpha$-attractors, primordial black holes and gravitational wave
  backgrounds},
  \href{https://doi.org/10.1088/1475-7516/2023/04/033}{\emph{Journal of
  Cosmology and Astroparticle Physics} {\bfseries 2023}{\bfseries (04)} (2023)
  033} [\href{https://arxiv.org/abs/2211.14262}{{\ttfamily 2211.14262}}].

\bibitem{Qin:2023lgo}
W.~Qin \textit{et~al}., \emph{Planck constraints and gravitational wave
  forecasts for primordial black hole dark matter seeded by multifield
  inflation}, \href{https://doi.org/10.1103/PhysRevD.108.043508}{\emph{Physical
  Review D} {\bfseries 108}{\bfseries (4)} (2023) 043508}
  [\href{https://arxiv.org/abs/2303.02168}{{\ttfamily 2303.02168}}].

\bibitem{Animali_2023}
C.~Animali and V.~Vennin, \emph{Primordial black holes from stochastic
  tunnelling},
  \href{https://doi.org/10.1088/1475-7516/2023/02/043}{\emph{Journal of
  Cosmology and Astroparticle Physics} {\bfseries 2023}{\bfseries (02)} (2023)
  043} [\href{https://arxiv.org/abs/2210.03812}{{\ttfamily 2210.03812}}].

\bibitem{KAMENSHCHIK2019201}
A.~Y. Kamenshchik, A.~Tronconi, T.~Vardanyan and G.~Venturi,
  \emph{Non-canonical inflation and primordial black holes production},
  \href{https://doi.org/10.1016/j.physletb.2019.02.036}{\emph{Physics Letters
  B} {\bfseries 791} (2019) 201}
  [\href{https://arxiv.org/abs/1812.02547}{{\ttfamily 1812.02547}}].

\bibitem{romano2020sound}
A.~E. Romano, \emph{Sound speed induced production of primordial black holes},
  (2020) [\href{https://arxiv.org/abs/2006.07321}{{\ttfamily 2006.07321}}].

\bibitem{Cai_2018}
Y.-F. Cai, X.~Tong, D.-G. Wang and S.-F. Yan, \emph{{Primordial Black Holes
  from Sound Speed Resonance during Inflation}},
  \href{https://doi.org/10.1103/physrevlett.121.081306}{\emph{Physical Review
  Letters} {\bfseries 121}{\bfseries (8)} (2018) 081306}
  [\href{https://arxiv.org/abs/1805.03639}{{\ttfamily 1805.03639}}].

\bibitem{Palma:2020_Seeding}
G.~A. Palma, S.~Sypsas and C.~Zenteno, \emph{{Seeding Primordial Black Holes in
  Multifield Inflation}},
  \href{https://doi.org/10.1103/PhysRevLett.125.121301}{\emph{Physical Review
  Letters} {\bfseries 125}{\bfseries (12)} (2020) 121301}
  [\href{https://arxiv.org/abs/2004.06106}{{\ttfamily 2004.06106}}].

\bibitem{Braglia:2020eai}
M.~Braglia \textit{et~al}., \emph{{Generating PBHs and small-scale GWs in
  two-field models of inflation}},
  \href{https://doi.org/10.1088/1475-7516/2020/08/001}{\emph{Journal of
  Cosmology and Astroparticle Physics} {\bfseries 2020}{\bfseries (08)} (2020)
  001} [\href{https://arxiv.org/abs/2005.02895}{{\ttfamily 2005.02895}}].

\bibitem{Braglia:2020taf}
M.~Braglia, X.~Chen and D.~K. Hazra, \emph{Probing primordial features with the
  stochastic gravitational wave background},
  \href{https://doi.org/10.1088/1475-7516/2021/03/005}{\emph{Journal of
  Cosmology and Astroparticle Physics} {\bfseries 2021}{\bfseries (03)} (2021)
  005} [\href{https://arxiv.org/abs/2012.05821}{{\ttfamily 2012.05821}}].

\bibitem{Fumagalli:2020_Turning}
J.~Fumagalli, S.~Renaux-Petel, J.~W. Ronayne and L.~T. Witkowski,
  \emph{{Turning in the landscape: A new mechanism for generating primordial
  black holes}},
  \href{https://doi.org/10.1016/j.physletb.2023.137921}{\emph{Physics Letters
  B} {\bfseries 841} (2023) 137921}
  [\href{https://arxiv.org/abs/2004.08369}{{\ttfamily 2004.08369}}].

\bibitem{Ahmed:2021ucx}
W.~Ahmed, M.~Junaid and U.~Zubair, \emph{Primordial black holes and
  gravitational waves in hybrid inflation with chaotic potentials},
  \href{https://doi.org/10.1016/j.nuclphysb.2022.115968}{\emph{Nuclear Physics
  B} {\bfseries 984} (2022) 115968}
  [\href{https://arxiv.org/abs/2109.14838}{{\ttfamily 2109.14838}}].

\bibitem{Geller:2022nkr}
S.~R. Geller, W.~Qin, E.~McDonough and D.~I. Kaiser, \emph{Primordial black
  holes from multifield inflation with nonminimal couplings},
  \href{https://doi.org/10.1103/PhysRevD.106.063535}{\emph{Physical Review D}
  {\bfseries 106}{\bfseries (6)} (2022) 063535}
  [\href{https://arxiv.org/abs/2205.04471}{{\ttfamily 2205.04471}}].

\bibitem{Iacconi_2022}
L.~Iacconi, H.~Assadullahi, M.~Fasiello and D.~Wands, \emph{Revisiting
  small-scale fluctuations in $\alpha$-attractor models of inflation},
  \href{https://doi.org/10.1088/1475-7516/2022/06/007}{\emph{Journal of
  Cosmology and Astroparticle Physics} {\bfseries 2022}{\bfseries (06)} (2022)
  007} [\href{https://arxiv.org/abs/2112.05092}{{\ttfamily 2112.05092}}].

\bibitem{Kawai:2022emp}
S.~Kawai and J.~Kim, \emph{Primordial black holes and gravitational waves from
  nonminimally coupled supergravity inflation},
  \href{https://doi.org/10.1103/PhysRevD.107.043523}{\emph{Physical Review D}
  {\bfseries 107}{\bfseries (4)} (2023) 043523}
  [\href{https://arxiv.org/abs/2209.15343}{{\ttfamily 2209.15343}}].

\bibitem{Ozsoy:2023ryl}
O.~Özsoy and G.~Tasinato, \emph{{Inflation and Primordial Black Holes}},
  \href{https://doi.org/10.3390/universe9050203}{\emph{Universe} {\bfseries
  9}{\bfseries (5)} (2023) 203}
  [\href{https://arxiv.org/abs/2301.03600}{{\ttfamily 2301.03600}}].

\bibitem{Mishra:2019pzq}
S.~S. Mishra and V.~Sahni, \emph{Primordial black holes from a tiny bump/dip in
  the inflaton potential},
  \href{https://doi.org/10.1088/1475-7516/2020/04/007}{\emph{Journal of
  Cosmology and Astroparticle Physics} {\bfseries 2020}{\bfseries (04)} (2020)
  007} [\href{https://arxiv.org/abs/1911.00057}{{\ttfamily 1911.00057}}].

\bibitem{ZhengRuiFeng:2021zoz}
R.~Zheng, J.~Shi and T.~Qiu, \emph{On primordial black holes and secondary
  gravitational waves generated from inflation with solo/multi-bumpy
  potential}, \href{https://doi.org/10.1088/1674-1137/ac42bd}{\emph{Chinese
  Physics C} {\bfseries 46}{\bfseries (4)} (2022) 045103}
  [\href{https://arxiv.org/abs/2106.04303}{{\ttfamily 2106.04303}}].

\bibitem{Inomata:2021tpx}
K.~Inomata, E.~McDonough and W.~Hu, \emph{Amplification of primordial
  perturbations from the rise or fall of the inflaton},
  \href{https://doi.org/10.1088/1475-7516/2022/02/031}{\emph{Journal of
  Cosmology and Astroparticle Physics} {\bfseries 2022}{\bfseries (02)} (2022)
  031} [\href{https://arxiv.org/abs/2110.14641}{{\ttfamily 2110.14641}}].

\bibitem{Frolovsky_2022}
D.~Frolovsky, S.~V. Ketov and S.~Saburov, \emph{E-models of inflation and
  primordial black holes},
  \href{https://doi.org/10.3389/fphy.2022.1005333}{\emph{Frontiers in Physics}
  {\bfseries 10} (2022) } [\href{https://arxiv.org/abs/2207.11878}{{\ttfamily
  2207.11878}}].

\bibitem{Inomata:2018epa}
K.~Inomata and T.~Nakama, \emph{Gravitational waves induced by scalar
  perturbations as probes of the small-scale primordial spectrum},
  \href{https://doi.org/10.1103/PhysRevD.99.043511}{\emph{Physical Review D}
  {\bfseries 99}{\bfseries (4)} (2019) 043511}
  [\href{https://arxiv.org/abs/1812.00674}{{\ttfamily 1812.00674}}].

\bibitem{Dalianis:2018ymb}
I.~Dalianis, \emph{Constraints on the curvature power spectrum from primordial
  black hole evaporation},
  \href{https://doi.org/10.1088/1475-7516/2019/08/032}{\emph{Journal of
  Cosmology and Astroparticle Physics} {\bfseries 2019}{\bfseries (08)} (2019)
  032} [\href{https://arxiv.org/abs/1812.09807}{{\ttfamily 1812.09807}}].

\bibitem{Kalaja:2019_From}
A.~Kalaja \textit{et~al}., \emph{From primordial black holes abundance to
  primordial curvature power spectrum (and back)},
  \href{https://doi.org/10.1088/1475-7516/2019/10/031}{\emph{Journal of
  Cosmology and Astroparticle Physics} {\bfseries 2019}{\bfseries (10)} (2019)
  031} [\href{https://arxiv.org/abs/1908.03596}{{\ttfamily 1908.03596}}].

\bibitem{Peskin:1995ev}
M.~E. Peskin and D.~V. Schroeder, \emph{{An Introduction to quantum field
  theory}}, Westview Press (1995).

\bibitem{Haymaker:1983xk}
R.~W. Haymaker and J.~Perez-Mercader, \emph{Convexity of the effective
  potential}, \href{https://doi.org/10.1103/PhysRevD.27.1948}{\emph{Physical
  Review D} {\bfseries 27}{\bfseries (8)} (1983) 1948}.

\bibitem{Amer:1982ri}
A.~Amer \textit{et~al}., \emph{Vacuum instability criterion in an effective
  potential approach},
  \href{https://doi.org/10.1016/0550-3213(83)90664-8}{\emph{Nuclear Physics B}
  {\bfseries 214}{\bfseries (2)} (1983) 299}.

\bibitem{Casas:1998cf}
J.~A. Casas, V.~Di~Clemente and M.~Quirós, \emph{The effective potential in
  the presence of several mass scales},
  \href{https://doi.org/10.1016/S0550-3213(99)00262-X}{\emph{Nuclear Physics B}
  {\bfseries 553}{\bfseries (1--2)} (1999) 511}
  [\href{https://arxiv.org/abs/hep-ph/9809275}{{\ttfamily hep-ph/9809275}}].

\bibitem{Achucarro:2012fd}
A.~Achúcarro, J.-O. Gong, G.~A. Palma and S.~P. Patil, \emph{Correlating
  features in the primordial spectra},
  \href{https://doi.org/10.1103/PhysRevD.87.121301}{\emph{Physical Review D}
  {\bfseries 87}{\bfseries (12)} (2013) 121301}
  [\href{https://arxiv.org/abs/1211.5619}{{\ttfamily 1211.5619}}].

\bibitem{Isidori:2007vm}
G.~Isidori, V.~S. Rychkov, A.~Strumia and N.~Tetradis, \emph{Gravitational
  corrections to standard model vacuum decay},
  \href{https://doi.org/10.1103/PhysRevD.77.025034}{\emph{Physical Review D}
  {\bfseries 77}{\bfseries (2)} (2008) 025034}
  [\href{https://arxiv.org/abs/0712.0242}{{\ttfamily 0712.0242}}].

\bibitem{Ragavendra:2020sop}
H.~V. Ragavendra, P.~Saha, L.~Sriramkumar and J.~Silk, \emph{Primordial black
  holes and secondary gravitational waves from ultraslow roll and punctuated
  inflation}, \href{https://doi.org/10.1103/PhysRevD.103.083510}{\emph{Physical
  Review D} {\bfseries 103}{\bfseries (8)} (2021) 083510}
  [\href{https://arxiv.org/abs/2008.12202}{{\ttfamily 2008.12202}}].

\bibitem{Bhaumik_2020}
N.~Bhaumik and R.~K. Jain, \emph{Primordial black holes dark matter from
  inflection point models of inflation and the effects of reheating},
  \href{https://doi.org/10.1088/1475-7516/2020/01/037}{\emph{Journal of
  Cosmology and Astroparticle Physics} {\bfseries 2020}{\bfseries (01)} (2020)
  037} [\href{https://arxiv.org/abs/1907.04125}{{\ttfamily 1907.04125}}].

\bibitem{Gangopadhyay_2022}
M.~R. Gangopadhyay, J.~C. Jain, D.~Sharma and Yogesh, \emph{Production of
  primordial black holes via single field inflation and observational
  constraints},
  \href{https://doi.org/10.1140/epjc/s10052-022-10796-x}{\emph{The European
  Physical Journal C} {\bfseries 82}{\bfseries (9)} (2022) 849}
  [\href{https://arxiv.org/abs/2108.13839}{{\ttfamily 2108.13839}}].

\bibitem{Germani17}
C.~Germani and T.~Prokopec, \emph{On primordial black holes from an inflection
  point}, \href{https://doi.org/10.1016/j.dark.2017.09.001}{\emph{Physics of
  the Dark Universe} {\bfseries 18} (2017) 6}
  [\href{https://arxiv.org/abs/1706.04226}{{\ttfamily 1706.04226}}].

\bibitem{Cicoli:2018asa}
M.~Cicoli, V.~A. Diaz and F.~G. Pedro, \emph{Primordial black holes from string
  inflation},
  \href{https://doi.org/10.1088/1475-7516/2018/06/034}{\emph{Journal of
  Cosmology and Astroparticle Physics} {\bfseries 2018}{\bfseries (06)} (2018)
  034} [\href{https://arxiv.org/abs/1803.02837}{{\ttfamily 1803.02837}}].

\bibitem{Ezquiaga:2017fvi}
J.~M. Ezquiaga, J.~García-Bellido and E.~Ruiz~Morales, \emph{{Primordial black
  hole production in Critical Higgs Inflation}},
  \href{https://doi.org/10.1016/j.physletb.2017.11.039}{\emph{Physics Letters
  B} {\bfseries 776} (2018) 345}
  [\href{https://arxiv.org/abs/1705.04861}{{\ttfamily 1705.04861}}].

\bibitem{Ballesteros:2020qam}
G.~Ballesteros, J.~Rey, M.~Taoso and A.~Urbano, \emph{Primordial black holes as
  dark matter and gravitational waves from single-field polynomial inflation},
  \href{https://doi.org/10.1088/1475-7516/2020/07/025}{\emph{Journal of
  Cosmology and Astroparticle Physics} {\bfseries 2020}{\bfseries (07)} (2020)
  025} [\href{https://arxiv.org/abs/2001.08220}{{\ttfamily 2001.08220}}].

\bibitem{Ballesteros:2017fsr}
G.~Ballesteros and M.~Taoso, \emph{Primordial black hole dark matter from
  single field inflation},
  \href{https://doi.org/10.1103/PhysRevD.97.023501}{\emph{Physical Review D}
  {\bfseries 97}{\bfseries (2)} (2018) 023501}
  [\href{https://arxiv.org/abs/1709.05565}{{\ttfamily 1709.05565}}].

\bibitem{Gow:2021_ACPS}
A.~D. Gow, C.~T. Byrnes, P.~S. Cole and S.~Young, \emph{{The power spectrum on
  small scales: robust constraints and comparing PBH methodologies}},
  \href{https://doi.org/10.1088/1475-7516/2021/02/002}{\emph{Journal of
  Cosmology and Astroparticle Physics} {\bfseries 2021}{\bfseries (02)} (2021)
  002} [\href{https://arxiv.org/abs/2008.03289}{{\ttfamily 2008.03289}}].

\bibitem{Ozsoy_2018}
O.~Özsoy, S.~Parameswaran, G.~Tasinato and I.~Zavala, \emph{Mechanisms for
  primordial black hole production in string theory},
  \href{https://doi.org/10.1088/1475-7516/2018/07/005}{\emph{Journal of
  Cosmology and Astroparticle Physics} {\bfseries 2018}{\bfseries (07)} (2018)
  005} [\href{https://arxiv.org/abs/1803.07626}{{\ttfamily 1803.07626}}].

\bibitem{Ozsoy_2021}
O.~Özsoy and Z.~Lalak, \emph{Primordial black holes as dark matter and
  gravitational waves from bumpy axion inflation},
  \href{https://doi.org/10.1088/1475-7516/2021/01/040}{\emph{Journal of
  Cosmology and Astroparticle Physics} {\bfseries 2021}{\bfseries (01)} (2021)
  040} [\href{https://arxiv.org/abs/2008.07549}{{\ttfamily 2008.07549}}].

\bibitem{Domenech:2021ztg}
G.~Domenech, \emph{{Scalar Induced Gravitational Waves Review}},
  \href{https://doi.org/10.3390/universe7110398}{\emph{Universe} {\bfseries
  7}{\bfseries (11)} (2021) 398}
  [\href{https://arxiv.org/abs/2109.01398}{{\ttfamily 2109.01398}}].

\bibitem{Kawasaki:2019mbl}
M.~Kawasaki and H.~Nakatsuka, \emph{Effect of nonlinearity between density and
  curvature perturbations on the primordial black hole formation},
  \href{https://doi.org/10.1103/PhysRevD.99.123501}{\emph{Physical Review D}
  {\bfseries 99}{\bfseries (12)} (2019) 123501}
  [\href{https://arxiv.org/abs/1903.02994}{{\ttfamily 1903.02994}}].

\bibitem{Young_2019}
S.~Young, I.~Musco and C.~T. Byrnes, \emph{Primordial black hole formation and
  abundance: contribution from the non-linear relation between the density and
  curvature perturbation},
  \href{https://doi.org/10.1088/1475-7516/2019/11/012}{\emph{Journal of
  Cosmology and Astroparticle Physics} {\bfseries 2019}{\bfseries (11)} (2019)
  012} [\href{https://arxiv.org/abs/1904.00984}{{\ttfamily 1904.00984}}].

\bibitem{DeLuca:2019qsy}
V.~De~Luca \textit{et~al}., \emph{{The ineludible non-Gaussianity of the
  primordial black hole abundance}},
  \href{https://doi.org/10.1088/1475-7516/2019/07/048}{\emph{Journal of
  Cosmology and Astroparticle Physics} {\bfseries 2019}{\bfseries (07)} (2019)
  048} [\href{https://arxiv.org/abs/1904.00970}{{\ttfamily 1904.00970}}].

\bibitem{1981SvA....25..406P}
A.~G. Polnarev and M.~Y. Khlopov, \emph{{Primordial Black Holes and the ERA of
  Superheavy Particle Dominance in the Early Universe}},
  \href{https://ui.adsabs.harvard.edu/abs/1981SvA....25..406P/abstract}{\emph{Soviet
  Astronomy} {\bfseries 25} (1981) 406}.

\bibitem{Harada:2016mhb}
T.~Harada \textit{et~al}., \emph{{PRIMORDIAL BLACK HOLE FORMATION IN THE
  MATTER-DOMINATED PHASE OF THE UNIVERSE}},
  \href{https://doi.org/10.3847/1538-4357/833/1/61}{\emph{The Astrophysical
  Journal} {\bfseries 833}{\bfseries (1)} (2016) 61}
  [\href{https://arxiv.org/abs/1609.01588}{{\ttfamily 1609.01588}}].

\bibitem{Carr:2017edp}
B.~Carr, T.~Tenkanen and V.~Vaskonen, \emph{Primordial black holes from
  inflaton and spectator field perturbations in a matter-dominated era},
  \href{https://doi.org/10.1103/PhysRevD.96.063507}{\emph{Physical Review D}
  {\bfseries 96}{\bfseries (6)} (2017) 063507}
  [\href{https://arxiv.org/abs/1706.03746}{{\ttfamily 1706.03746}}].

\bibitem{Cole:2017gle}
P.~S. Cole and C.~T. Byrnes, \emph{Extreme scenarios: the tightest possible
  constraints on the power spectrum due to primordial black holes},
  \href{https://doi.org/10.1088/1475-7516/2018/02/019}{\emph{Journal of
  Cosmology and Astroparticle Physics} {\bfseries 2018}{\bfseries (02)} (2018)
  019} [\href{https://arxiv.org/abs/1706.10288}{{\ttfamily 1706.10288}}].

\bibitem{Allahverdi:2020bys}
R.~Allahverdi \textit{et~al}., \emph{{The First Three Seconds: a Review of
  Possible Expansion Histories of the Early Universe}},
  \href{https://doi.org/10.21105/astro.2006.16182}{\emph{The Open Journal of
  Astrophysics} (2020) } [\href{https://arxiv.org/abs/2006.16182}{{\ttfamily
  2006.16182}}].

\bibitem{Dienes:2021woi}
K.~R. Dienes \textit{et~al}., \emph{{Stasis in an expanding universe: A recipe
  for stable mixed-component cosmological eras}},
  \href{https://doi.org/10.1103/PhysRevD.105.023530}{\emph{Physical Review D}
  {\bfseries 105}{\bfseries (2)} (2022) 023530}
  [\href{https://arxiv.org/abs/2111.04753}{{\ttfamily 2111.04753}}].

\bibitem{Ballesteros_2020}
G.~Ballesteros, J.~Rey and F.~Rompineve, \emph{Detuning primordial black hole
  dark matter with early matter domination and axion monodromy},
  \href{https://doi.org/10.1088/1475-7516/2020/06/014}{\emph{Journal of
  Cosmology and Astroparticle Physics} {\bfseries 2020}{\bfseries (06)} (2020)
  014} [\href{https://arxiv.org/abs/1912.01638}{{\ttfamily 1912.01638}}].

\bibitem{Bullock:1996_Non-Gaussian}
J.~S. Bullock and J.~R. Primack, \emph{{Non-Gaussian fluctuations and
  primordial black holes from inflation}},
  \href{https://doi.org/10.1103/PhysRevD.55.7423}{\emph{Physical Review D}
  {\bfseries 55}{\bfseries (12)} (1997) 7423}
  [\href{https://arxiv.org/abs/astro-ph/9611106}{{\ttfamily
  astro-ph/9611106}}].

\bibitem{Young:2015_Influence}
S.~Young, D.~Regan and C.~T. Byrnes, \emph{Influence of large local and
  non-local bispectra on primordial black hole abundance},
  \href{https://doi.org/10.1088/1475-7516/2016/02/029}{\emph{Journal of
  Cosmology and Astroparticle Physics} {\bfseries 2016}{\bfseries (02)} (2016)
  029} [\href{https://arxiv.org/abs/1512.07224}{{\ttfamily 1512.07224}}].

\bibitem{Yoo:2019_Abundance}
C.-M. Yoo, J.-O. Gong and S.~Yokoyama, \emph{{Abundance of primordial black
  holes with local non-Gaussianity in peak theory}},
  \href{https://doi.org/10.1088/1475-7516/2019/09/033}{\emph{Journal of
  Cosmology and Astroparticle Physics} {\bfseries 2019}{\bfseries (09)} (2019)
  033} [\href{https://arxiv.org/abs/1906.06790}{{\ttfamily 1906.06790}}].

\bibitem{Palma:2020_Non-Gaussian}
G.~A. Palma, B.~Scheihing~Hitschfeld and S.~Sypsas, \emph{{Non-Gaussian CMB and
  LSS statistics beyond polyspectra}},
  \href{https://doi.org/10.1088/1475-7516/2020/02/027}{\emph{Journal of
  Cosmology and Astroparticle Physics} {\bfseries 2020}{\bfseries (02)} (2020)
  027} [\href{https://arxiv.org/abs/1907.05332}{{\ttfamily 1907.05332}}].

\bibitem{Taoso:2021_Non-gaussianities}
M.~Taoso and A.~Urbano, \emph{Non-gaussianities for primordial black hole
  formation},
  \href{https://doi.org/10.1088/1475-7516/2021/08/016}{\emph{Journal of
  Cosmology and Astroparticle Physics} {\bfseries 2021}{\bfseries (08)} (2021)
  016} [\href{https://arxiv.org/abs/2102.03610}{{\ttfamily 2102.03610}}].

\bibitem{Young:2022_non-G}
S.~Young, \emph{Peaks and primordial black holes: the effect of
  non-gaussianity},
  \href{https://doi.org/10.1088/1475-7516/2022/05/037}{\emph{Journal of
  Cosmology and Astroparticle Physics} {\bfseries 2022}{\bfseries (05)} (2022)
  037} [\href{https://arxiv.org/abs/2201.13345}{{\ttfamily 2201.13345}}].

\bibitem{Fujita:2013_Algorithm}
T.~Fujita, M.~Kawasaki, Y.~Tada and T.~Takesako, \emph{A new algorithm for
  calculating the curvature perturbations in stochastic inflation},
  \href{https://doi.org/10.1088/1475-7516/2013/12/036}{\emph{Journal of
  Cosmology and Astroparticle Physics} {\bfseries 2013}{\bfseries (12)} (2013)
  036} [\href{https://arxiv.org/abs/1308.4754}{{\ttfamily 1308.4754}}].

\bibitem{Vennin:2015_Correlation}
V.~Vennin and A.~A. Starobinsky, \emph{Correlation functions in stochastic
  inflation}, \href{https://doi.org/10.1140/epjc/s10052-015-3643-y}{\emph{The
  European Physical Journal C} {\bfseries 75} (2015) 413}
  [\href{https://arxiv.org/abs/1506.04732}{{\ttfamily 1506.04732}}].

\bibitem{Pattison:2017_Quantum}
C.~Pattison, V.~Vennin, H.~Assadullahi and D.~Wands, \emph{Quantum diffusion
  during inflation and primordial black holes},
  \href{https://doi.org/10.1088/1475-7516/2017/10/046}{\emph{Journal of
  Cosmology and Astroparticle Physics} {\bfseries 2017}{\bfseries (10)} (2017)
  046} [\href{https://arxiv.org/abs/1707.00537}{{\ttfamily 1707.00537}}].

\bibitem{Ezquiaga:2019_Exponential}
J.~M. Ezquiaga, J.~García-Bellido and V.~Vennin, \emph{The exponential tail of
  inflationary fluctuations: consequences for primordial black holes},
  \href{https://doi.org/10.1088/1475-7516/2020/03/029}{\emph{Journal of
  Cosmology and Astroparticle Physics} {\bfseries 2020}{\bfseries (03)} (2020)
  029} [\href{https://arxiv.org/abs/1912.05399}{{\ttfamily 1912.05399}}].

\bibitem{Figueroa:2020_Non-Gaussian}
D.~G. Figueroa, S.~Raatikainen, S.~Räsänen and E.~Tomberg,
  \emph{{Non-Gaussian Tail of the Curvature Perturbation in Stochastic
  Ultraslow-Roll Inflation: Implications for Primordial Black Hole
  Production}},
  \href{https://doi.org/10.1103/PhysRevLett.127.101302}{\emph{Physical Review
  Letters} {\bfseries 127}{\bfseries (10)} (2021) 101302}
  [\href{https://arxiv.org/abs/2012.06551}{{\ttfamily 2012.06551}}].

\bibitem{Ando:2020_Power}
K.~Ando and V.~Vennin, \emph{Power spectrum in stochastic inflation},
  \href{https://doi.org/10.1088/1475-7516/2021/04/057}{\emph{Journal of
  Cosmology and Astroparticle Physics} {\bfseries 2021}{\bfseries (04)} (2021)
  057} [\href{https://arxiv.org/abs/2012.02031}{{\ttfamily 2012.02031}}].

\bibitem{Pattison:2021_USR}
C.~Pattison, V.~Vennin, D.~Wands and H.~Assadullahi, \emph{Ultra-slow-roll
  inflation with quantum diffusion},
  \href{https://doi.org/10.1088/1475-7516/2021/04/080}{\emph{Journal of
  Cosmology and Astroparticle Physics} {\bfseries 2021}{\bfseries (04)} (2021)
  080} [\href{https://arxiv.org/abs/2101.05741}{{\ttfamily 2101.05741}}].

\bibitem{Rigopoulos:2021_Inflation}
G.~Rigopoulos and A.~Wilkins, \emph{{Inflation is always semi-classical:
  diffusion domination overproduces Primordial Black Holes}},
  \href{https://doi.org/10.1088/1475-7516/2021/12/027}{\emph{Journal of
  Cosmology and Astroparticle Physics} {\bfseries 2021}{\bfseries (12)} (2021)
  027} [\href{https://arxiv.org/abs/2107.05317}{{\ttfamily 2107.05317}}].

\bibitem{Tada:2021_Statistics}
Y.~Tada and V.~Vennin, \emph{Statistics of coarse-grained cosmological fields
  in stochastic inflation},
  \href{https://doi.org/10.1088/1475-7516/2022/02/021}{\emph{Journal of
  Cosmology and Astroparticle Physics} {\bfseries 2022}{\bfseries (02)} (2022)
  021} [\href{https://arxiv.org/abs/2111.15280}{{\ttfamily 2111.15280}}].

\bibitem{Jackson:2022_Numerical}
J.~H.~P. Jackson \textit{et~al}., \emph{Numerical simulations of stochastic
  inflation using importance sampling},
  \href{https://doi.org/10.1088/1475-7516/2022/10/067}{\emph{Journal of
  Cosmology and Astroparticle Physics} {\bfseries 2022}{\bfseries (10)} (2022)
  067} [\href{https://arxiv.org/abs/2206.11234}{{\ttfamily 2206.11234}}].

\bibitem{Gow:2022jfb}
A.~D. Gow \textit{et~al}., \emph{{Non-perturbative non-Gaussianity and
  primordial black holes}},
  \href{https://doi.org/10.1209/0295-5075/acd417}{\emph{Europhysics Letters}
  {\bfseries 142}{\bfseries (4)} (2023) 49001}
  [\href{https://arxiv.org/abs/2211.08348}{{\ttfamily 2211.08348}}].

\bibitem{Ferrante:2022mui}
G.~Ferrante, G.~Franciolini, A.~J. Iovino and A.~Urbano, \emph{{Primordial
  non-Gaussianity up to all orders: Theoretical aspects and implications for
  primordial black hole models}},
  \href{https://doi.org/10.1103/PhysRevD.107.043520}{\emph{Physical Review D}
  {\bfseries 107}{\bfseries (4)} (2023) 043520}
  [\href{https://arxiv.org/abs/2211.01728}{{\ttfamily 2211.01728}}].

\bibitem{Nakama:2016kfq}
T.~Nakama, T.~Suyama and J.~Yokoyama, \emph{Supermassive black holes formed by
  direct collapse of inflationary perturbations},
  \href{https://doi.org/10.1103/PhysRevD.94.103522}{\emph{Physical Review D}
  {\bfseries 94}{\bfseries (10)} (2016) 103522}
  [\href{https://arxiv.org/abs/1609.02245}{{\ttfamily 1609.02245}}].

\bibitem{Nakama:2019htb}
T.~Nakama, K.~Kohri and J.~Silk, \emph{Ultracompact minihalos associated with
  stellar-mass primordial black holes},
  \href{https://doi.org/10.1103/PhysRevD.99.123530}{\emph{Physical Review D}
  {\bfseries 99}{\bfseries (12)} (2019) 123530}
  [\href{https://arxiv.org/abs/1905.04477}{{\ttfamily 1905.04477}}].

\bibitem{Tada:2015noa}
Y.~Tada and S.~Yokoyama, \emph{Primordial black holes as biased tracers},
  \href{https://doi.org/10.1103/PhysRevD.91.123534}{\emph{Physical Review D}
  {\bfseries 91}{\bfseries (12)} (2015) 123534}
  [\href{https://arxiv.org/abs/1502.01124}{{\ttfamily 1502.01124}}].

\bibitem{Young:2014oea}
S.~Young and C.~T. Byrnes, \emph{Long-short wavelength mode coupling tightens
  primordial black hole constraints},
  \href{https://doi.org/10.1103/PhysRevD.91.083521}{\emph{Physical Review D}
  {\bfseries 91}{\bfseries (8)} (2015) 083521}
  [\href{https://arxiv.org/abs/1411.4620}{{\ttfamily 1411.4620}}].

\bibitem{vanLaak:2023ppj}
R.~van Laak and S.~Young, \emph{{Primordial black hole isocurvature modes from
  non-Gaussianity}},  (2023)
  [\href{https://arxiv.org/abs/2303.05248}{{\ttfamily 2303.05248}}].

\bibitem{Passaglia_2019}
S.~Passaglia, W.~Hu and H.~Motohashi, \emph{{Primordial black holes and local
  non-Gaussianity in canonical inflation}},
  \href{https://doi.org/10.1103/physrevd.99.043536}{\emph{Physical Review D}
  {\bfseries 99}{\bfseries (4)} (2019) 043536}
  [\href{https://arxiv.org/abs/1812.08243}{{\ttfamily 1812.08243}}].

\bibitem{Ballesteros:2020sre}
G.~Ballesteros, J.~Rey, M.~Taoso and A.~Urbano, \emph{Stochastic inflationary
  dynamics beyond slow-roll and consequences for primordial black hole
  formation},
  \href{https://doi.org/10.1088/1475-7516/2020/08/043}{\emph{Journal of
  Cosmology and Astroparticle Physics} {\bfseries 2020}{\bfseries (08)} (2020)
  043} [\href{https://arxiv.org/abs/2006.14597}{{\ttfamily 2006.14597}}].

\bibitem{Planck:2018jri}
{\scshape Planck} collaboration, \emph{{Planck 2018 results. X. Constraints on
  inflation}},
  \href{https://doi.org/10.1051/0004-6361/201833887}{\emph{Astronomy \&
  Astrophysics} {\bfseries 641} (2020) A10}
  [\href{https://arxiv.org/abs/1807.06211}{{\ttfamily 1807.06211}}].

\bibitem{Tristram:2021tvh}
M.~Tristram \textit{et~al}., \emph{{Improved limits on the tensor-to-scalar
  ratio using BICEP and Planck data}},
  \href{https://doi.org/10.1103/PhysRevD.105.083524}{\emph{Physical Review D}
  {\bfseries 105}{\bfseries (8)} (2022) 083524}
  [\href{https://arxiv.org/abs/2112.07961}{{\ttfamily 2112.07961}}].

\bibitem{Ashoorioon:2020hln}
A.~Ashoorioon, A.~Rostami and J.~T. Firouzjaee, \emph{Examining the end of
  inflation with primordial black holes mass distribution and gravitational
  waves}, \href{https://doi.org/10.1103/PhysRevD.103.123512}{\emph{Physical
  Review D} {\bfseries 103}{\bfseries (12)} (2021) 123512}
  [\href{https://arxiv.org/abs/2012.02817}{{\ttfamily 2012.02817}}].

\bibitem{Kawai:2021edk}
S.~Kawai and J.~Kim, \emph{{Primordial black holes from Gauss-Bonnet-corrected
  single field inflation}},
  \href{https://doi.org/10.1103/PhysRevD.104.083545}{\emph{Physical Review D}
  {\bfseries 104}{\bfseries (8)} (2021) 083545}
  [\href{https://arxiv.org/abs/2108.01340}{{\ttfamily 2108.01340}}].

\bibitem{Talebian:2022cwk}
A.~Talebian, S.~A. Hosseini~Mansoori and H.~Firouzjahi, \emph{{Inflation from
  Multiple Pseudo-scalar Fields: Primordial Black Hole Dark Matter and
  Gravitational Waves}},
  \href{https://doi.org/10.3847/1538-4357/acc8d2}{\emph{The Astrophysical
  Journal} {\bfseries 948}{\bfseries (1)} (2023) 48}
  [\href{https://arxiv.org/abs/2210.13822}{{\ttfamily 2210.13822}}].

\bibitem{Papanikolaou:2022did}
T.~Papanikolaou, A.~Lymperis, S.~Lola and E.~N. Saridakis, \emph{Primordial
  black holes and gravitational waves from non-canonical inflation},
  \href{https://doi.org/10.1088/1475-7516/2023/03/003}{\emph{Journal of
  Cosmology and Astroparticle Physics} {\bfseries 2023}{\bfseries (03)} (2023)
  003} [\href{https://arxiv.org/abs/2211.14900}{{\ttfamily 2211.14900}}].

\bibitem{Martin:2019nuw}
J.~Martin, T.~Papanikolaou and V.~Vennin, \emph{Primordial black holes from the
  preheating instability in single-field inflation},
  \href{https://doi.org/10.1088/1475-7516/2020/01/024}{\emph{Journal of
  Cosmology and Astroparticle Physics} {\bfseries 2020}{\bfseries (01)} (2020)
  024} [\href{https://arxiv.org/abs/1907.04236}{{\ttfamily 1907.04236}}].

\bibitem{Dvali:2021byy}
G.~Dvali, F.~Kühnel and M.~Zantedeschi, \emph{Primordial black holes from
  confinement},
  \href{https://doi.org/10.1103/PhysRevD.104.123507}{\emph{Physical Review D}
  {\bfseries 104}{\bfseries (12)} (2021) 123507}
  [\href{https://arxiv.org/abs/2108.09471}{{\ttfamily 2108.09471}}].

\bibitem{Coleman:1969sm}
S.~Coleman, J.~Wess and B.~Zumino, \emph{{Structure of phenomenological
  Lagrangians. I.}},
  \href{https://doi.org/10.1103/PhysRev.177.2239}{\emph{Physical Review}
  {\bfseries 177}{\bfseries (5)} (1969) 2239}.

\bibitem{Burgess:2007pt}
C.~P. Burgess, \emph{{An Introduction to Effective Field Theory}},
  \href{https://doi.org/10.1146/annurev.nucl.56.080805.140508}{\emph{Annual
  Review of Nuclear and Particle Science} {\bfseries 57} (2007) 329}
  [\href{https://arxiv.org/abs/hep-th/0701053}{{\ttfamily hep-th/0701053}}].

\bibitem{Brivio:2017vri}
I.~Brivio and M.~Trott, \emph{The standard model as an effective field theory},
  \href{https://doi.org/10.1016/j.physrep.2018.11.002}{\emph{Physics Reports}
  {\bfseries 793} (2019) 1} [\href{https://arxiv.org/abs/1706.08945}{{\ttfamily
  1706.08945}}].

\bibitem{Simpson:2017qvj}
F.~Simpson, R.~Jimenez, C.~Pena-Garay and L.~Verde, \emph{{Strong Bayesian
  evidence for the normal neutrino hierarchy}},
  \href{https://doi.org/10.1088/1475-7516/2017/06/029}{\emph{Journal of
  Cosmology and Astroparticle Physics} {\bfseries 2017}{\bfseries (06)} (2017)
  029} [\href{https://arxiv.org/abs/1703.03425}{{\ttfamily 1703.03425}}].

\bibitem{Schwetz:2017fey}
T.~Schwetz \textit{et~al}., \emph{{Comment on ``Strong Evidence for the Normal
  Neutrino Hierarchy''}},  (2017)
  [\href{https://arxiv.org/abs/1703.04585}{{\ttfamily 1703.04585}}].

\bibitem{Franciolini:2018vbk}
G.~Franciolini, A.~Kehagias, S.~Matarrese and A.~Riotto, \emph{{Primordial
  black holes from inflation and non-Gaussianity}},
  \href{https://doi.org/10.1088/1475-7516/2018/03/016}{\emph{Journal of
  Cosmology and Astroparticle Physics} {\bfseries 2018}{\bfseries (03)} (2018)
  016} [\href{https://arxiv.org/abs/1801.09415}{{\ttfamily 1801.09415}}].

\bibitem{Starobinsky:1992ts}
A.~A. Starobinsky, \emph{Spectrum of adiabatic perturbations in the universe
  when there are singularities in the inflation potential},
  \href{http://jetpletters.ru/ps/1276/article_19291.shtml}{\emph{Journal of
  Experimental and Theoretical Physics Letters} {\bfseries 55}{\bfseries (9)}
  (1992) 489}.

\bibitem{Pi:2022zxs}
S.~Pi and J.~Wang, \emph{{Primordial Black Hole Formation in Starobinsky's
  Linear Potential Model}},  (2022)
  [\href{https://arxiv.org/abs/2209.14183}{{\ttfamily 2209.14183}}].

\end{thebibliography}\endgroup

\end{document}